\documentclass[aps,prl,twocolumn,showpacs,superscriptaddress]{revtex4-1}
\usepackage{amsmath}
\usepackage{epsfig}
\usepackage{color}
\usepackage{endnotes}
\usepackage{natbib}
\usepackage[colorlinks,breaklinks]{hyperref}
\usepackage{nicefrac}
\usepackage{bm}
\usepackage{dcolumn}
\usepackage{graphics}
\usepackage{graphicx} 
\usepackage{tikz} 
\usepackage{amsmath}
\usepackage{amssymb}
\usepackage{color}
\usepackage{changes}
\usepackage{multirow}
\usepackage{lineno}

\usepackage{titlesec}
\titlespacing*{\section}{0pt}{0.8\baselineskip}{0.8\baselineskip}

\renewcommand{\figurename}{\bf Fig.}

\newcommand*{\MIT}{Massachusetts Institute of Technology, Cambridge, Massachusetts 02139, USA}
\newcommand*{\TAU}{School of Physics and Astronomy, Tel Aviv University, Tel Aviv 69978, Israel}
\newcommand*{\TUD}{Institut f\"ur Kernphysik, Technische Universit\"at Darmstadt, 64289 Darmstadt, Germany}
\newcommand*{\GSI}{GSI Helmholtzzentrum f\"ur Schwerionenforschung GmbH, Planckstr. 1, 64291 Darmstadt, Germany}
\newcommand*{\FAIR}{Helmholtz Forschungsakademie Hessen f\"ur FAIR, Max-von-Laue-Str. 12, 60438 Frankfurt, Germany}
\newcommand*{\JINR}{Joint Institute for Nuclear Research, Dubna 141980, Russia}
\newcommand*{\CEA}{IRFU, CEA, Universit\'{e} Paris-Saclay, F-91191 Gif-sur-Yvette, France}
\newcommand*{\DSU}{Dubna State University, Dubna 141980, Russia}

\begin{document}

\def\Label{}


   \title{Unperturbed inverse kinematics nucleon knockout measurements with a 48 GeV/c carbon beam}

\author{M. Patsyuk}
\affiliation{\MIT}
\affiliation{\JINR}

\author{J. Kahlbow}
\affiliation{\MIT}
\affiliation{\TAU}

\author{G. Laskaris}
\affiliation{\MIT}
\affiliation{\TAU}

\author{M. Duer}
\affiliation{\TUD}

\author{V. Lenivenko}
\affiliation{\JINR}

\author{E.P. Segarra}
\affiliation{\MIT}

\author{T. Atovullaev}
\affiliation{\JINR}
\affiliation{\DSU}

\author{G. Johansson}
\affiliation{\TAU}

\author{T. Aumann}
\affiliation{\TUD}
\affiliation{\GSI}
\affiliation{\FAIR}
\author{A. Corsi}
\affiliation{\CEA}
\author{O. Hen}
\email[Contact Author \ ]{(hen@mit.edu)}
\affiliation{\MIT}
\author{M. Kapishin}
\affiliation{\JINR}
\author{V. Panin}
\affiliation{\CEA}
\affiliation{\GSI}
\author{E. Piasetzky$^3$ et al.}

\collaboration{The BM@N Collaboration}

\maketitle

{\bf 
  From superconductors to atomic nuclei, strongly-interacting
  many-body systems are ubiquitous in nature. 
  Measuring the microscopic structure of such systems is a
  formidable challenge, often met by particle knockout scattering experiments~\cite{JACOB:1966zz,kelly96}.
  While such measurements are fundamental for mapping the structure of atomic nuclei
\cite{kelly96,Gade:2008zz,Kobayashi:2008zzc,Wakasa:2017rsk,duer18},
  their interpretation is often challenged by quantum mechanical 
  initial- and final-state interactions (ISI/FSI) of the incoming 
  and scattered particles~\cite{JACOB:1966zz,Hansen:2003sn,kelly96,Cosyn:2009bi,Atti:2015eda}.  
  Here we overcome this fundamental limitation by measuring the quasi-free scattering of 48 GeV/c $^{12}$C
  ions from hydrogen. 
  The distribution of single protons is studied by detecting two protons 
  at large angles in coincidence with an intact $^{11}$B nucleus.
  The $^{11}$B detection is shown to select the transparent part of the reaction and exclude the otherwise
  large ISI/FSI that would break the $^{11}$B apart.
  By further detecting residual $^{10}$B and $^{10}$Be nuclei, we also
  identified short-range correlated (SRC) nucleon-nucleon pairs~\cite{subedi08,Feldmeier:2011qy,hen14,Hen:2016kwk,Atti:2015eda}, 
  and provide direct experimental evidence for the separation of the pair wave-function from that of the residual many-body nuclear system~\cite{Atti:2015eda,Cruz-Torres:2019fum}.
  All measured reactions are well described by theoretical calculations that do not contain ISI/FSI distortions.
  Our results thus showcase a new ability to study the short-distance
  structure of short-lived radioactive atomic nuclei at the forthcoming
  FAIR~\cite{Spiller:2006gj} and FRIB~\cite{FRIB400} facilities. 
  These studies will be pivotal for developing a
  ground-breaking microscopic understanding of the structure and properties
  of nuclei far from stability and the formation of visible matter in the universe.}

\begin{figure*}[t]
\centering  
	\includegraphics[width=0.5\linewidth]{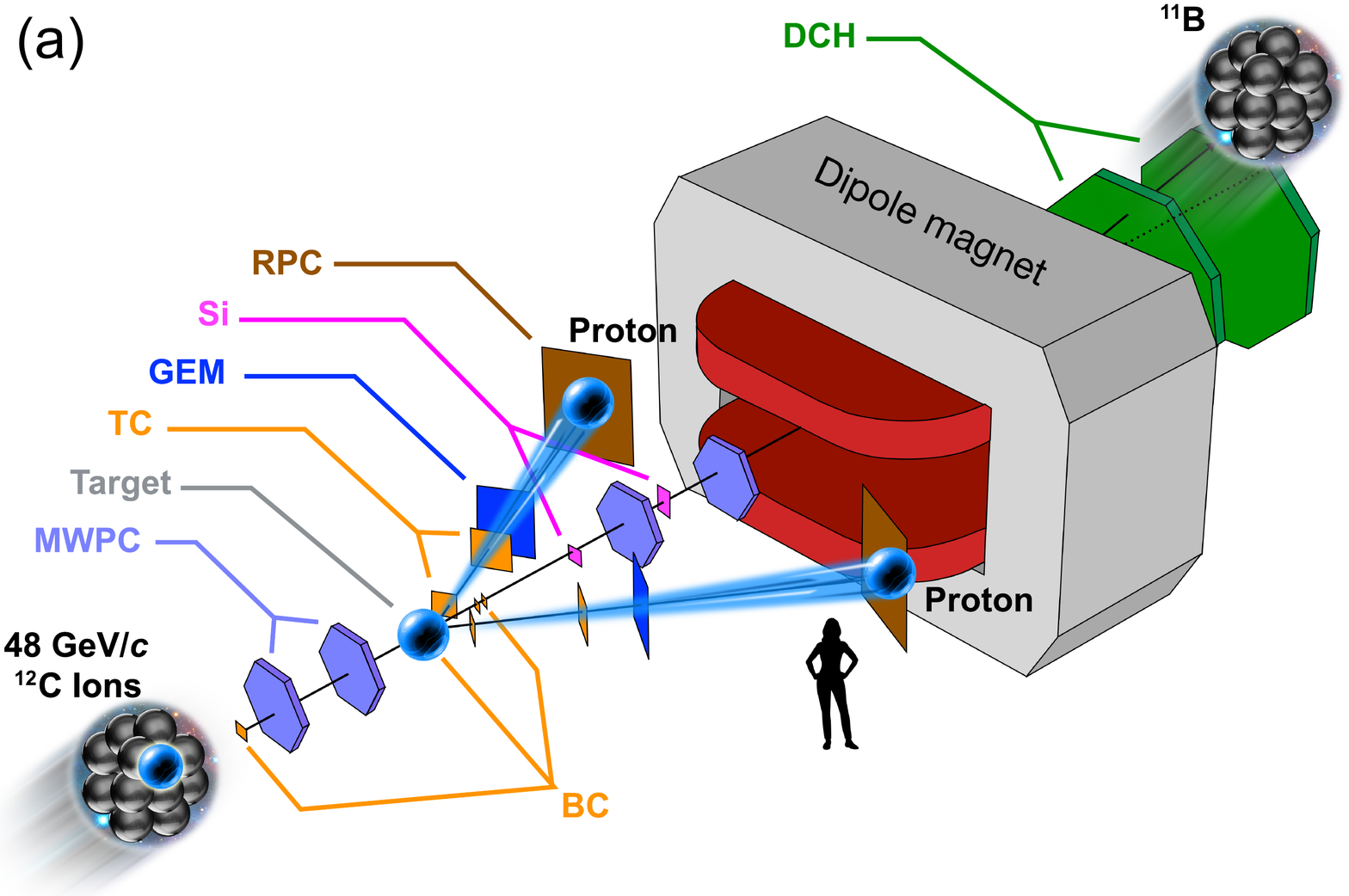}\hspace*{0.8cm}
	\includegraphics[width=0.365\linewidth]{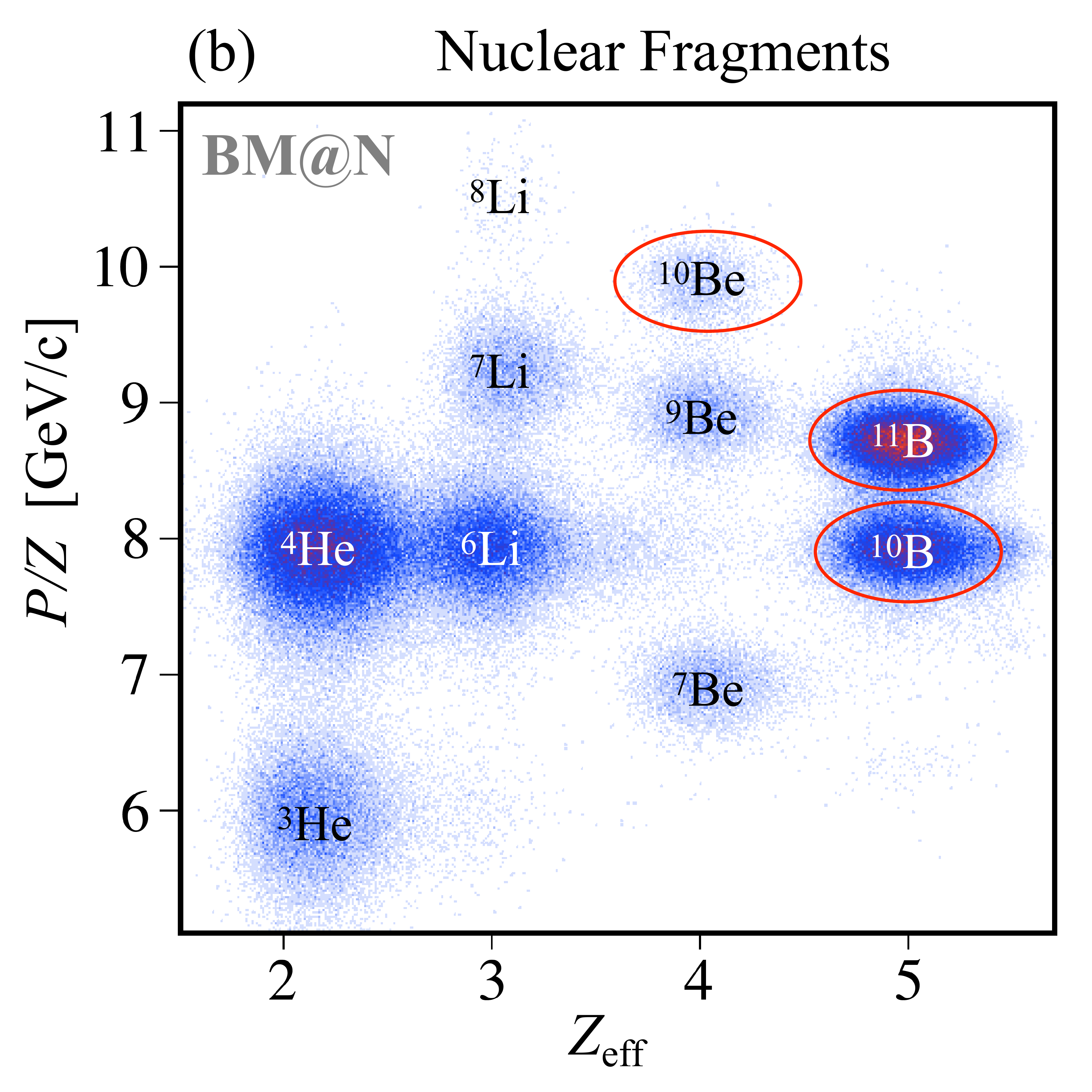}	
	\caption{{\bf $\vert$ Experimental setup and fragment identification.} 
	{\normalfont (a) Carbon nuclei traveling at $48$ GeV/c hit protons in a liquid hydrogen target, knocking out individual protons from the beam-ion. Position- and time-sensitive detectors (MWPC, GEM, RPC, Si, and DCH) are used to track the incoming ion beam, knockout protons, and residual nuclear fragments and determine their momenta. (b) The bend of the nuclear fragments in the large dipole magnet, combined with charge measurements with the beam counters (BC) allows identifying the various fragments.
	In this work we refer to events with detected $^{11}$B, $^{10}$B, and $^{10}$Be heavy fragments, see text for details.}}	
\label{fig:setup}
\end{figure*}

Strongly-interacting systems are difficult to study.
In the special case of strongly-interacting quantum gasses,
ground-state properties can be directly measured using ultra-cold atomic traps, 
where one can instantaneously turn off the interactions between the atoms and the trap itself~\cite{Mukherjee_PRL}.
This allows exploring a wide range of fundamental quantum mechanical phenomena 
and to imitate strongly correlated states in condensed matter systems where 
similar control over inter-particle interactions cannot be obtained~\cite{Bloch_RMP}.

Due to their high-density and complex strong interaction,
constructing such model systems for atomic nuclei is extremely challenging.
Instead, the distribution of nucleons in nuclei is traditionally studied using high-energy
electron scattering experiments that detect the scattered electron and knockout nucleon 
with high-resolution spectrometers~\cite{kelly96}.
ISI/FSI cause a reduction of the quasielastic cross-section (attenuation) as well as distortion of the reconstructed single nucleon ground state properties.
Pre-selection of the reaction kinematics or post-selection of the un-detected residual nucleus
allows suppressing ISI/FSI distortions and use energy and momentum conservation to reconstruct the 
distribution of nucleons in the nucleus~\cite{kelly96,Atti:2015eda,Hen:2016kwk,schmidt20,Cruz-Torres:2020uke}.

While largely limited to stable nuclei, such experiments helped establish the 
nuclear shell model~\cite{kelly96} and the existence and dynamics of SRC nucleon
pairs~\cite{subedi08,hen14,duer18,schmidt20} that constitute the next significant 
approximation to nuclear structure after the shell model~\cite{kelly96,Hen:2016kwk,Atti:2015eda}.

Extending these studies to radioactive nuclei far from nuclear stability is a growing frontier of nuclear science~\cite{Hansen:2003sn}.
Such studies require performing scattering experiments in inverse kinematics, 
where low luminosity high-energy beams of radioactive nuclei are scattered from protons in hydrogen targets~\cite{Obertelli:2011ck}.
The cross-section for such reactions is significantly higher than that for electron scattering, 
but comes at the price of large ISI that prevents kinematical pre-selection.
Additionally, since there is rarely sufficient energy resolution to determine the residual 
nuclear state from the measured momenta of the knocked-out nucleons, 
post-selection requires direct detection of the residual nuclear system~\cite{Atar:2018dhg}.

Here we use post-selection in high-energy inverse kinematics $(p,2p)$ scattering to probe 
single-particle states and SRCs in the well understood $^{12}$C nucleus. 
By detecting a bound nuclear fragment we select the transparent part of the
scattering reaction where single step scattering dominates 
and distortions due to ISI/FSI of the incoming/outgoing nucleons are suppressed.

By identifying $^{11}$B fragment we successfully study the distribution of protons 
in the $p$-shell of $^{12}$C, where we obtain consistent distributions 
for both quasielastic (QE) and inelastic (IE) scattering reactions. 
Selecting $^{10}$B and $^{10}$Be fragments we further identify, 
for the first time in inverse kinematics, the hard breakup of SRC pairs.
We  directly measure the pair motion in the nucleus and establish the separation 
of the strong inter-pair interaction from the residual nuclear system.
The latter is a key feature of modern theoretical SRC models~\cite{Feldmeier:2011qy,Atti:2015eda,Hen:2016kwk,Chen:2016bde,Cruz-Torres:2019fum}, 
that has not been experimentally confirmed.

While significantly reducing the measured event rate, these post-selection requirements  
are shown to ensure that the measured reaction has little to no sensitivity to ISI/FSI induced distortions,
thereby opening the door to studying the single-particle and short-distance structure of nuclei far from stability.

\section{Experimental setup}
The experiment took place at the Joint Institute for Nuclear Research (JINR), using a 4~GeV/c/nucleon ion beam from the Nuclotron accelerator, a stationary liquid-hydrogen target, and a modified BM@N 
(Baryonic Matter at Nuclotron) experimental setup, as shown in Fig.~\ref{fig:setup}a.

The beam was monitored upstream the target using thin scintillator-based beam counters (BCs) used for charge identification, 
a veto counter (V-BC) for beam-halo rejection, and two multi-wire proportional chambers (MWPCs) for event-by-event beam tracking.
The BC closer to the target was also used to define the event start time $t_0$. 

A two-arm spectrometer (TAS) was placed downstream of the target to detect the 
two protons from the $(p,2p)$ reaction that emerge between $24^\circ$ and $37^\circ$,
corresponding to $90^\circ$ QE scattering in the two-protons center-of-mass (c.m) frame. 
Each spectrometer arm consisted of two scintillator trigger counters (TC), a gas electron multiplier (GEM) station and a multi-gap resistive plate chamber (RPC) wall.

Proton tracks were reconstructed using their hit location in the GEM and RPC walls.
We only consider events where the distance of closest approach between the proton tracks is smaller than $4$~cm and the interaction vertex of each proton is reconstructed within the central $26$~cm of the target (Extended Data Fig.~\ref{fig:lh2_vertex}). 
The time difference between the RPC and $t_0$ signals define the proton time of flight (TOF), 
that is used to determine its momentum from the measured track length, assuming a proton mass.

As the protons of interest for our analysis have momenta between $1.5$ and $2.5$~GeV/c ($0.85 < \beta < 0.935$),
we conservatively reject events with proton tracks having $\beta > 0.96$ or $ < 0.8$.

Signals from the TC were combined with the BCs upstream the  target 
to form the main $^{12}$C$(p,2p)$ reaction trigger for the experiment. 
Additional triggers were set up for monitoring and calibration purposes, 
see Online Supplementary Materials for details.

Nuclear fragments following the $(p,2p)$ reaction are emitted at small angles 
with respect to the incident beam with momentum, that is similar to the per nucleon beam momentum.
Three silicon (Si) planes and two MWPCs were placed in the beam-line downstream the target to measure the fragment scattering angle.
Following the MWPCs the fragments enter a large acceptance $2.87$ T$\cdot$m dipole magnet.
Two drift chambers (DCH) are used to measure the fragment trajectory after the magnet.

The fragment momenta are determined from their measured trajectories through the the dipole magnet.
Fragments are identified from the combination of their rigidity ($P/Z$) in the magnet and energy deposition
in the two scintillator BCs placed between the target and the magnet entrance, see Fig.~\ref{fig:setup}b. 
The latter is proportional to the sum of all fragment charges squared ($Z_{\mathrm{eff}} = \sqrt{\sum Z^2}$).

See Methods and Online Supplementary Materials for additional details on the experimental setup and data analysis procedures.

\begin{figure}[t]
\centering  
	\includegraphics[width=\linewidth]{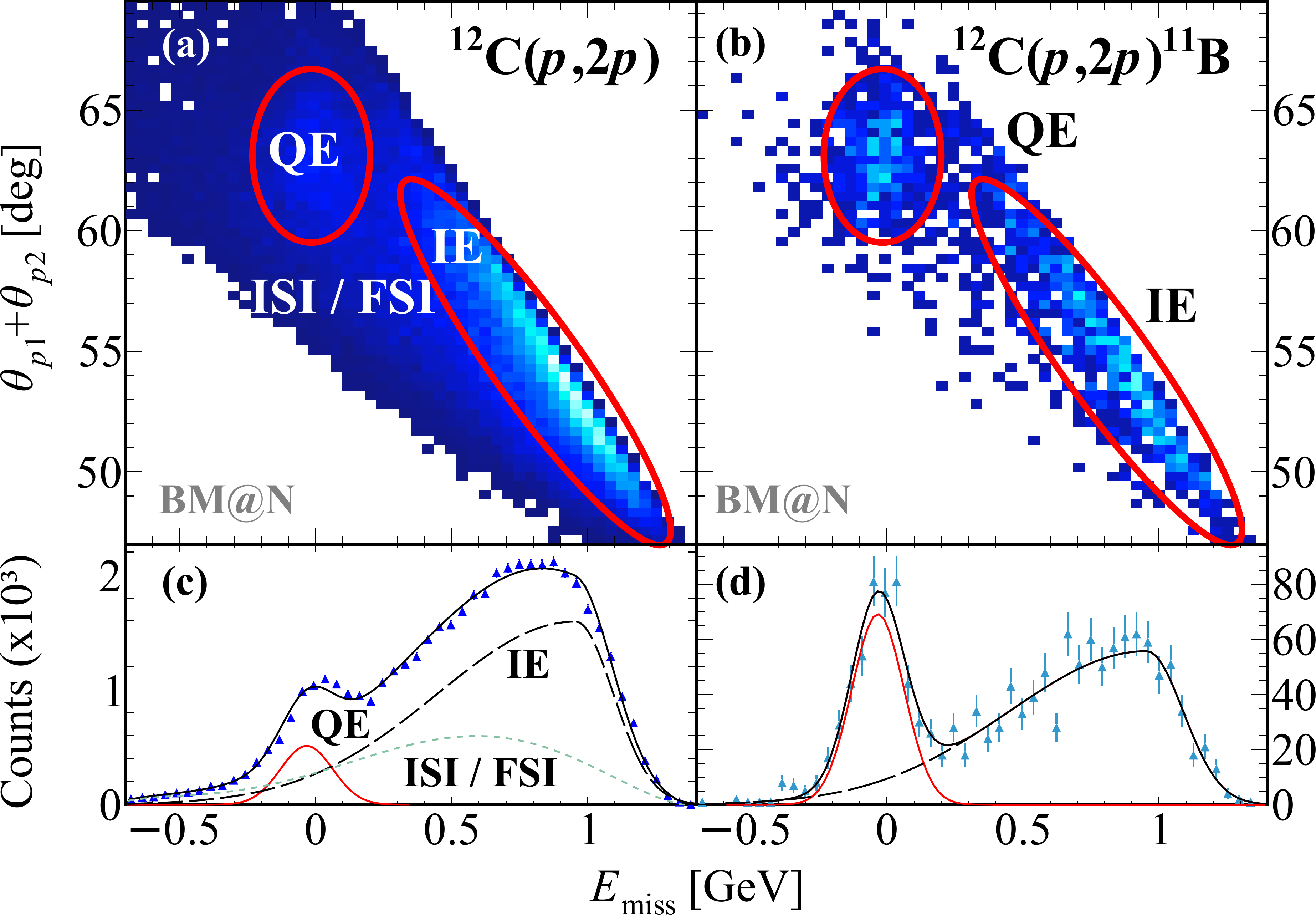}
	\caption{{\bf $\vert$ Quasi-Free Scattering (QFS) distributions.} 
	{\normalfont (a) and (b): The correlation between the measured missing-energy ($E_{\mathrm{miss}}$, calculated in the $^{12}$C rest-frame) and the measured lab-frame two-proton in-plane opening angle ($\theta_1+\theta_2$). Distributions are shown for inclusive $^{12}$C$(p,2p)$ events (a) and exclusive $^{12}$C$(p,2p)^{11}$B events (b). (c) and (d): One-dimensional projections of the missing-energy distributions for inclusive (c) and exclusive (d) events (see Extended Data Fig. 2a for opening angle projections). Data error bars show statistical uncertainties of the data at the 1$\sigma$ confidence level. The lines show the results of a fit to the measured quasielastic (QE) and inelastic (IE) peaks, using the same functional form for both distributions. The inclusive distribution requires an additional fit component, associated with ISI/FSI distortions, to fully describe the data. QE events are seen as a peak around low missing energy and opening angles of $\sim63^\circ$. IE reactions populate higher missing-energy and lower opening angles while ISI/FSI populate both regions and the ridge between them in the inclusive spectra.  }}
\label{fig:MF_Em_Pm}
\end{figure}

\section{Single proton knockout}
We identify exclusive $^{12}$C$(p,2p)^{11}$B events by requiring the detection of a 
$^{11}$B fragment in coincidence with two charged particle tracks in the TAS.
Energy and momentum conservation for this reaction reads:
\begin{equation}
	{\bar p}_{^{12}\mathrm{C}} + {\bar p}_{tg} = {\bar p}_1 + {\bar p}_2 +  {\bar p}_{^{11}\mathrm{B}},
\label{eq:1}
\end{equation}
where ${\bar p}_{^{12}\mathrm{C}}=(\sqrt{({\bf p}_{^{12}\mathrm{C}}^2+m_{^{12}\mathrm{C}}^2)},0,0,p_{^{12}\mathrm{C}})$ and ${\bar p}_{tg}=(m_p,0,0,0)$ are respectively the incident beam-ion and target proton four-momentum vectors.
${\bar p}_1$, ${\bar p}_2$, and ${\bar p}_{^{11}\mathrm{B}}$ are the four-momentum vectors of the detected protons and $^{11}$B fragment.
Assuming QE scattering off a nucleon which is moving in a mean-field potential, we can approximate ${\bar p}_{^{12}\mathrm{C}} = {\bar p}_{i} + {\bar p}_{^{11}\mathrm{B}}$, where ${\bar p}_{i}$ is the initial proton four-momentum inside the $^{12}$C ion.
Substituting into Eq.~\ref{eq:1} we obtain:
\begin{equation}
	{\bar p}_{i} \approx {\bar p}_{\mathrm{miss}} \equiv {\bar p}_1 + {\bar p}_2 - {\bar p}_{tg},
\label{eq:Mmiss_MF_p2p}
\end{equation}
where ${\bar p}_{\mathrm{miss}}$ is the measured missing four-momentum of the reaction and is only equal to ${\bar p}_{i}$ in the case of unperturbed (no ISI/FSI) QE scattering.
Through the text, the missing momentum vector is shown and discussed after being boosted from the lab-frame to the incident $^{12}$C ion rest-frame.

Figure~\ref{fig:MF_Em_Pm} shows the measured missing energy $E_{\mathrm{miss}}\equiv m_p-e_{\mathrm{miss}}$ (where $e_{\mathrm{miss}}$ is the energy component of ${\bar p}_{\mathrm{miss}}$ in the $^{12}$C rest-frame) distribution and its correlation with the lab-frame two-proton in-plane opening angle, $\theta_1+\theta_2$, for inclusive $^{12}$C$(p,2p)$ (left panels) and exclusive $^{12}$C$(p,2p)^{11}$B (right panels) events. 
Both distributions show two distinct regions:
(A) low missing-energy and large in-plane opening angles that correspond to QE scattering and
(B) high missing energy and small  in-plane opening angles  that correspond to IE scattering.

As seen in the $E_{\mathrm{miss}}$ projections, the inclusive $^{12}$C$(p,2p)$ events are contaminated by ISI/FSI backgrounds around and underlying both IE and QE regions.
This background is not evident in the $^{12}$C$(p,2p)^{11}$B case,
which is our first indication that requiring the coincidence detection of $^{11}$B fragments selects a
unique subset of one-step processes where a single nucleon was knocked-out without any further interaction
with the residual fragment.
We note  that while bound excited states cannot be separated from the ground state in $^{12}$C$(p,2p)^{11}$B events, their contribution  is small~\cite{Panin:2016div} and should not impact the measured  momentum distribution. See Methods for details.

\begin{figure*}[t]
\centering  
	\includegraphics[width=0.8\linewidth]{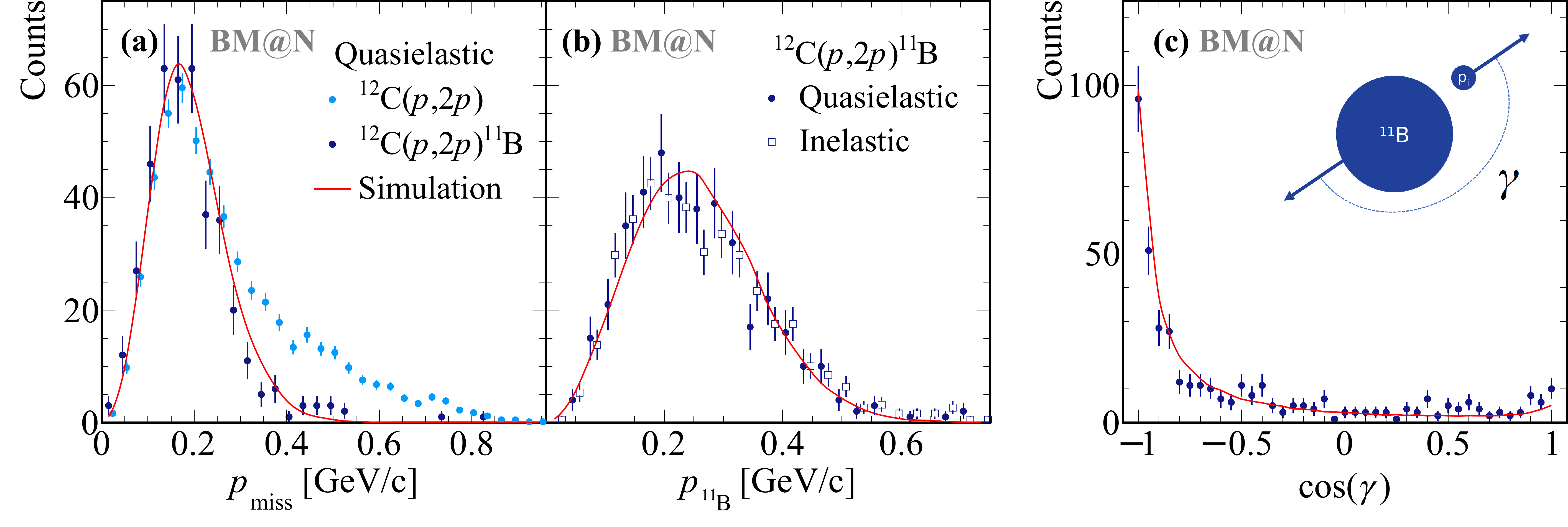}
	\caption{{\bf $\vert$ Momentum distributions and angular correlation.} 
	{\normalfont (a) Missing-momentum distribution in $^{12}$C rest-frame for quasielastic $^{12}$C$(p,2p)$ and $^{12}$C$(p,2p)^{11}$B events.  
	(b) $^{11}$B fragment momentum distribution in $^{12}$C rest-frame for quasielastic and inelastic $^{12}$C$(p,2p)^{11}$B events. 
	The light blue points in (a) and the open symbols in (b) have a small artificial offset for better visibility. 
	(c) Distribution of the cosine of the opening-angle between the missing- and fragment-momentum in the plane transverse to the beam. 
	Solid red line shows the result of our quasielastic reaction simulation. 
	Data error bars show statistical uncertainties at the $1\sigma$ confidence level.
	The y-axis shows the counts for the quasielastic distribution. 
	The inelastic distributions are normalized to the peak region of the quasielastic distribution.
	All variables are shown in the $^{12}$C rest-frame.}}
\label{fig:MF_Distributions}
\end{figure*}

Figure~\ref{fig:MF_Distributions}a shows further evidence for  ISI/FSI suppression by comparing the measured missing-momentum distribution for $^{12}$C$(p,2p)$ QE events with and without $^{11}$B  tagging. The QE selection was done using the missing-energy and in-plane opening-angle cuts depicted in Fig.~\ref{fig:MF_Em_Pm} following a $2\sigma$ selection (see Methods for details).
The measured $^{12}$C$(p,2p)$ QE events show a significant high-momentum tail
that extends well beyond the nuclear Fermi-momentum ($\approx250$ MeV/c)
and is characteristic for ISI/FSI~\cite{Atti:2015eda}. This tail is completely suppressed by the $^{11}$B detection.

Figure~\ref{fig:MF_Distributions}b compares the measured $^{11}$B momentum 
distribution in the $^{12}$C rest-frame for both QE and IE $^{12}$C$(p,2p)^{11}$B events.
The fragment momentum distribution is equal for both reactions.
This shows that the observation of  a bound  fragment selects quasi-free unperturbed single-step reactions,
even in the case of hard inelastic $NN$ scattering and in a kinematical region which is otherwise dominated by FSI events.

In true unperturbed single-step $^{12}$C$(p,2p)^{11}$B QE scattering the measured missing- and fragment-momenta should balance each other. Fig.~\ref{fig:MF_Distributions}c shows the distribution of the cosine of the opening angle between the missing- and fragment-momenta in the plane transverse to the incident beam-ion (which is insensitive to boost effects and is measured with better resolution). While broadened due to detector resolutions,
a clear back-to-back correlation is observed which is a distinct signature of QE reactions.

The data shown in Fig.~\ref{fig:MF_Distributions} are compared to theoretical calculations 
of QE $(p,2p)$ scattering off a $p$-shell nucleon in $^{12}$C.
The calculation is implemented via a simulation that accounts for the experimental acceptance and detector resolutions,
uses measured $^1$H$(p,2p)$ elastic scattering cross section, and does not include ISI/FSI effects.
The total simulated event yield was scaled to match the data. See Methods for details.
The calculation agrees well with all measured $^{12}$C$(p,2p)^{11}$B distributions,
including the fragment momentum distribution for IE events
and the distribution of the angle between the missing- and fragment-momenta 
(including its tail induced by detector-resolution).

Additional data-theory comparisons are shown in Extended Data Fig.~\ref{fig:supp_angPP} and~\ref{fig:supp_pm} exhibiting good agreement.
This is a clear indication that the  $^{11}$B detection strongly suppresses
ISI/FSI, providing access to ground-state properties of $^{12}$C.

Comparing the tagged and inclusive reaction yields we find that in $\sim 50\%$ of the measured inclusive $^{12}$C$(p,2p)$ QE reactions
the residual nucleus is fragmented to lighter fragments ($Z<4$).
Specifically for heavier fragments, the $^{12}$C$(p,2p)^{11}$B QE event yield accounts for $(43.7\pm 2.4\,(\mathrm{stat})^{+4.9}_{-5.8}\,(\mathrm{sys}))\%$ of the measured inclusive $^{12}$C$(p,2p)$ QE yield,
and $^{12}$C$(p,2p)^{10}$B and  $^{12}$C$(p,2p)^{10}$Be QE events, due to QE scattering to an excited $^{11}$B state that de-excites via nucleon emission, account for an additional $(7.8\pm 1.0\,(\mathrm{stat})^{+1.3}_{-1.4}\,(\mathrm{sys}))\%$ and $\le 2\%$, respectively.
See Methods for details.

\section{Hard breakup of SRC pairs}
Next we study SRCs by measuring the $^{12}$C$(p,2p)^{10}$B and $^{12}$C$(p,2p)^{10}$Be reactions.
SRC breakup reactions produce $^{10}$B and $^{10}$Be fragments when interacting with a proton-neutron ($pn$) or proton-proton ($pp$) pair, respectively.
The fragment selection guarantees exclusion of secondary scattering processes, as shown above,
and restricts the excitation-energy of the residual A-2 system to below its nucleon separation energy. 
Furthermore, the fragment detection offers a direct experimental probe for the interaction between the SRC pair nucleons and the residual $A-2$ nucleons.

While $^{10}$B and $^{10}$Be fragments can be produced in SRC breakup reaction, they can also be produced following $(p,2p)$ interactions involving mean-field nucleons. 
As discussed above, $\sim 10\%$ of the measured inclusive mean-field $^{12}$C$(p,2p)$ QE events produce excited $^{11}$B fragment that decay to $^{10}$B and $^{10}$Be via nucleon emission.
These processes can be suppressed by requiring $|{\bf p}_{\mathrm{miss}}| > 350$~MeV/c,
which selects protons with initial momenta that is well above the nuclear Fermi level where SRCs predominate over mean-field nucleons~\cite{Hen:2016kwk}. See Methods for details.

High ${\bf p}_{\mathrm{miss}}$ $^{12}$C$(p,2p)^{10}$B and $^{12}$C$(p,2p)^{10}$Be events can also result from IE interactions that produce additional particles. 
Such reactions can involve mean-field nucleons and will not be suppressed by the high ${\bf p}_{\mathrm{miss}}$ requirement.
However, as shown in Fig.~\ref{fig:MF_Em_Pm}, they can be suppressed by restricting the missing-energy of the reaction and requiring a large in-plane opening angle between the measured $(p,2p)$ protons.

To guide this selection we used the Generalized Contact Formalism (GCF)~\cite{Cruz-Torres:2019fum}
to simulate $(p,2p)$ scattering events off SRC pairs (see Methods for details).
Following these calculations we select SRC breakup reactions by requiring an in-plane opening angle larger than $\sim63^\circ$ and
$-110 \leq E_{\mathrm{miss}} \leq 240$~MeV (see Extended Data Fig.~\ref{fig:Em_Pm_cut}).
We further use total-energy and momentum conservation to ensure exclusivity and suppress IE contributions by requiring a missing nucleon mass in the entire reaction: $ M_{\mathrm{miss,\:excl.}}^2 = ({\bar p}_{^{12}\mathrm{C}} + {\bar p}_{tg} - {\bar p}_1 - {\bar p}_2 -  {\bar p}_{^{10}\mathrm{B (Be)}})^2 \approx m_N^2$ (see Extended Data Fig.~\ref{fig:10B_Mm_exc}).

\begin{figure}[t]
\centering  
	\includegraphics[width=0.8\linewidth]{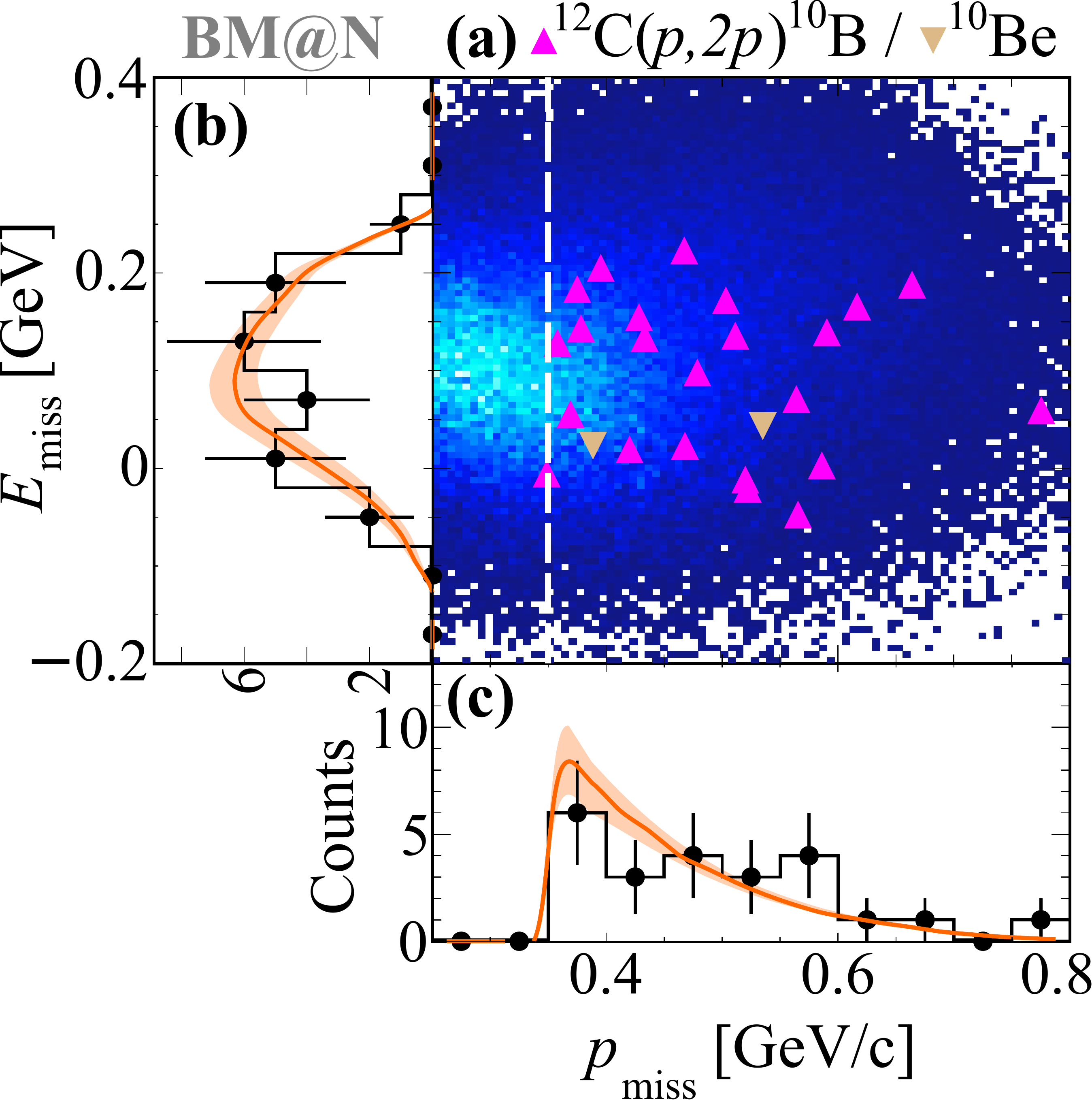}
	\caption{{\bf $\vert$ SRC  Selection in missing momentum and energy.}
	{\normalfont (a) Correlation between the missing-energy and missing-momentum for the measured $^{12}$C$(p,2p)^{10}$B (upwards facing purple triangles) and $^{12}$C$(p,2p)^{10}$Be (Downwards facing brown triangles) selected SRC events, on top of the GCF simulation (color scale). (b) and (c): One-dimensional projections for the measured (black points) and GCF simulated (orange line) missing-energy (b) and missing-momentum (c). The width of the bands and the data error bars show the systematic uncertainties of the model and the statistical uncertainties of the data, respectively, each at the 1$\sigma$ confidence level.}}
\label{fig:SRC_EmPm}
\end{figure}

Applying these selection cuts, we measured $23$ $^{12}$C$(p,2p)^{10}$B and $2$ $^{12}$C$(p,2p)^{10}$Be events.
The large $^{10}$B to $^{10}$Be event yield ratio is generally consistent with the previously observed predominance of $pn$- over $pp$-SRC pairs~\cite{piasetzky06,subedi08,hen14,Hen:2016kwk,Duer:2018sxh},
and is in full agreement with the GCF calculated $^{10}$B / $^{10}$Be yield ratio of $12.1$ obtained using input from ab-initio many-body calculations~\cite{Cruz-Torres:2019fum}. 
The observed $^{10}$B dominance also contradicts an expectation of similar $^{10}$B and $^{10}$Be yields if the measured reactions were dominated by mean field QE scattering followed by FSI with a single nucleon in $^{11}$B.

Figure~\ref{fig:SRC_EmPm} shows the missing-energy and missing-momentum distributions of the selected SRC $^{12}$C$(p,2p)^{10}$B events. The measured distributions show good agreement with the GCF predictions.
Additional kinematical distributions are shown and compared with the GCF in Extended Data Fig.~\ref{fig:10B_Pm} and~\ref{fig:10B_openAng_Pmn}. We specifically note that the distributions of the $z$-component of the missing-momentum is not centered around zero and is shifted towards the incident beam-direction (Extended Data Fig.~\ref{fig:10B_Pm}c). This is expected given the strong $s$-dependence of the
large-angle elementary proton-proton elastic scattering cross-section. See discussion in Methods.

\begin{figure}[t]
\centering  
	\includegraphics[width=\linewidth]{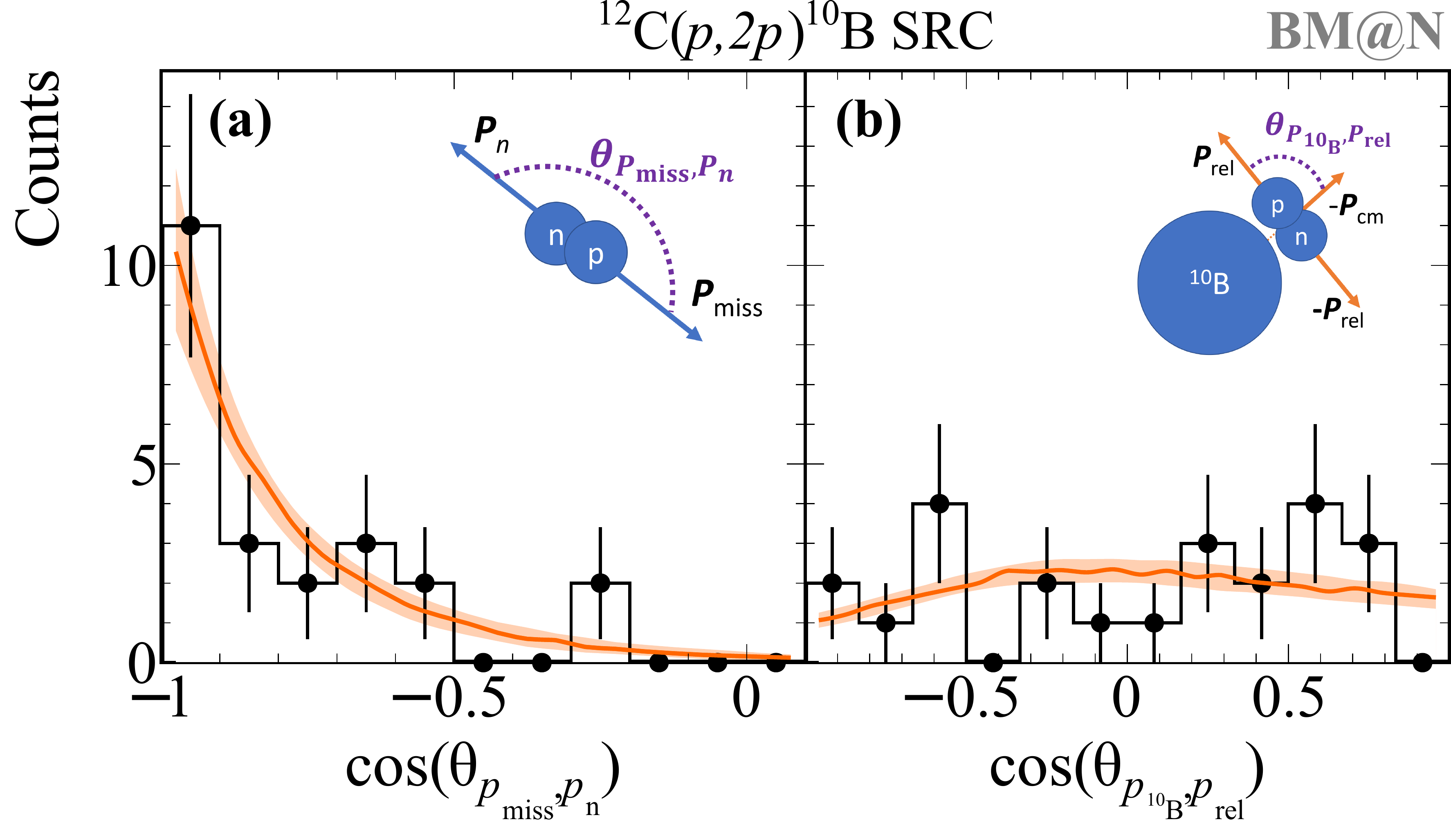}
	\caption{{\bf $\vert$ Angular correlations in SRC breakup events.}
	{\normalfont Distributions of the cosine of the angle between (a) the recoil nucleon and missing momentum and (b) $^{10}$B fragment and pair relative-momentum. Data (black points) are compared with GCF predictions (orange lines). The width of the bands and the data error bars show the systematic uncertainties of the model and the statistical uncertainties of the data, respectively, each at the 1$\sigma$ confidence level.}}
\label{fig:SRC_angles}
\end{figure}

Next we examine the angular correlations between the nucleons in the pair and between the pair and the $^{10}$B fragment.
Figure~\ref{fig:SRC_angles}a shows the distribution of the cosine of the angle between the missing momentum (Eq.~\ref{eq:Mmiss_MF_p2p}) and the reconstructed undetected recoil neutron momentum. 
A clear back-to-back correlation is observed, as expected for strongly-correlated nucleon pairs.
The width of the distribution is driven by the pair c.m. motion and detection resolutions.
It shows good agreement with the GCF prediction that assumes a three-dimensional Gaussian c.m. momentum distribution~\cite{Cruz-Torres:2019fum,Cohen:2018gzh}.

An independent determination of the SRC pair c.m. momentum distribution can be obtained from the $^{10}$B momentum distribution that is measured here for the first time (Extended Data Fig.~\ref{fig:10B_Pm}e-h).
We extract from the data an SRC pair c.m. momentum distribution with Gaussian width of $\sigma_{\mathrm{c.m.}}=(156\pm27)$~MeV/c (see Methods for extraction details), in agreement with previous electron scattering measurements~\cite{Cohen:2018gzh}.

Last we examine the factorization of the measured SRC pairs from the the residual nuclear system.
The strong two-body interaction between the nucleons in the pair was predicted~\cite{Atti:2015eda,Chen:2016bde,Cruz-Torres:2019fum} to allow modeling its distribution as independent functions of the pair relative and c.m. motion, with no correlation between them. Such factorization dramatically simplifies SRC calculations and should be evident experimentally by a lack of correlation between the pair c.m. and relative momenta.

Figure~\ref{fig:SRC_angles}b shows the distribution of the cosine of the angle between the 
$^{10}$B fragment momentum (i.\,e. pair c.m. momentum) and the pair relative momentum given by ${\bf p}_{\mathrm{rel}} =  ({\bf p}_{\mathrm{miss}} - {\bf p}_{n})/2$, where ${\bf p}_{n}$ is the reconstructed recoil neutron momenta.
The GCF assumes the above mentioned  factorization and therefore predicts a flat distribution,
that is slightly shaped by the acceptance of our detectors.
The data is in full agreement with this assumption. 

Therefore, by reporting here on the first measurement of SRC pairs with the detection of the residual bound $A-2$
nucleons system 
we are able to provide first direct experimental evidence for the factorization of SRC pairs from the many-body nuclear medium.

\section{Conclusions}
The dominant contributions of ISI/FSI to nucleon-knockout scattering measurements has  been a major difficulty
for experimentally extracting nucleon distributions in nuclei~\cite{Atti:2015eda,Hen:2016kwk,Bobeldijk:1994zz,Blomqvist:1998gq,benmokhtar05}.
Even in high-energy electron scattering at selected kinematics that minimize their contributions,
the remaining FSI effect had to be taken into account using theoretical estimates that introduce
significant model dependence to the obtained 
results~\cite{benmokhtar05,frankfurt08b,Atti:2015eda,Hen:2016kwk}.

At lower beam energies, the method of quasi-free proton-induced nucleon knockout in inverse kinematics has been recently developed and applied to study the single-particle structure of exotic nuclei~\cite{Kobayashi:2008zzc, Panin:2016div,Wakasa:2017rsk,Atar:2018dhg}. The data analysis and interpretation of these results heavily relies on the assumption that the extracted particle distributions are free from FSI contamination that has not been experimentally proven to date. 

Our findings however clearly demonstrate the feasibility of accessing properties of single-nucleons and SRC nucleon pairs in short-lived nuclei, in particular neutron-rich nuclei, using high-energy radioactive beams, produced at upcoming accelerator facilities such as FRIB and FAIR. 
With this method, we accomplished a big step towards realizing the goal of such facilities, which is exploring the formation of visible matter in the universe in the laboratory. The presented experimental method thus provides a basis to approximate, as closely as possible, the dense cold neutron-rich matter in neutron stars in the laboratory.\\

\bigbreak 
{\noindent \bf Acknowledgments} We acknowledge the efforts of the staff of the Accelerator and High-Energy Physics Divisions at JINR that made this experiment possible and I. Tserruya for fruitful discussions of the analysis and results.
We thank Yu. Borzunov, A. Konstantinov, D. Klimanskiy, I. Arkharov for providing stable operation of the Liquid Hydrogen cryogenic target.
The research was supported by the Israel Science Foundation, the Pazi Foundation, by the BMBF via Project No. 05P15RDFN1, through the GSI-TU Darmstadt co-operation agreement, by the U.S. DOE under grant No. DE- FG02-08ER41533 and by the Deutsche Forschungsgemeinschaft (DFG, German Research Foundation), Project-ID 279384907, SFB 1245, and the RFBR under grant numbers 18-02-40046, 20-38-90275, and 18-02-40084\textbackslash19.

\bigbreak 
{\noindent \bf Author Contributions} 
The experimental set up at the Nuclotron was designed and constructed by the BM@N Collaboration at JINR. 
Data reconstruction and calibration, Monte Carlo simulations of the detector and data analyses were performed by a
large number of BM@N Collaboration members, 
who also discussed and approved the scientific results.
In particular, the design and construction of the TAS was lead by G.L., who also led the data taking period.
The development and operation of the Data acquisition and trigger systems were lead by S.B. and V.Y., respectively.
The development and operation of the GEM and Silicon detectors were lead bt A.M. and N.Z., respectively.
Raw data processing and online monitoring were performed by S.M. and I.G.
M.R. contributed to the RPC analysis, V.P. contributed to the Si/MWPC analysis, D.B. contributed to the GEM analysis, and N.V. contributed to the DCH analysis.
The main data analysis was done by J.K., M.P., V.L., E.P.S., T.A., G.J., V.P., and M.D., 
with input from O.H., E.P., T.A., M.K. and A.C., and reviewed by the BM@N collaboration.\\

{\noindent \bf Competing interests} The authors declare no competing interests. \\

{\noindent \bf Data Availability} Source data are available for this paper. All other data that support the plots within this paper and other findings of this study are available from the corresponding author upon reasonable request. \\

{\noindent \bf Author Information} Reprints and permissions information is available at www.nature.com/reprints. Readers are welcome to comment on the online version of the paper. Publisher's note: Springer Nature remains neutral with regard to jurisdictional claims in published maps and institutional affiliations. Correspondence and requests for materials should be addressed to O.H. (hen@mit.edu).

\bigbreak 
{\noindent \bf Full Author List \\}
{M. Patsyuk},$^{	1	,	2			}$
{J. Kahlbow},$^{	1	,	3			}$
{G. Laskaris},$^{	1	,	3			}$
{M. Duer},$^{	4					}$
{V. Lenivenko},$^{	2					}$
{E.P. Segarra},$^{	1					}$
{T. Atovullaev},$^{	2	,	5			}$
{G. Johansson},$^{	3					}$
{T. Aumann},$^{	4	,	6	,	7	}$
{A. Corsi},$^{	8					}$
{O. Hen},$^{	1					}$
{M. Kapishin},$^{	2					}$
{V. Panin},$^{	8	,	6			}$
{E. Piasetzky},$^{	3					}$
{Kh. Abraamyan},$^{	2					}$
{S. Afanasiev},$^{	2					}$
{G. Agakishiev},$^{	2					}$
{P. Alekseev},$^{	9					}$
{I. Arkharov},$^{	21					}$
{E. Atkin},$^{	10					}$
{T. Aushev},$^{	11					}$
{V. Babkin},$^{	2					}$
{V. Balandin},$^{	2					}$
{D. Baranov},$^{	2					}$
{N. Barbashina},$^{	10					}$
{P. Batyuk},$^{	2					}$
{S. Bazylev},$^{	2					}$
{A. Beck},$^{	1					}$
{C.A. Bertulani},$^{	12					}$
{D. Blaschke},$^{	13					}$
{D. Blau},$^{	14					}$
{D. Bogoslovsky},$^{	2					}$
{A. Bolozdynya},$^{	10					}$
{K. Boretzky},$^{	6					}$
{Y. Borzunov,$^{	2					}$
{V. Burtsev},$^{	2					}$
{M. Buryakov},$^{	2					}$
{S. Buzin},$^{	2					}$
{A. Chebotov},$^{	2					}$
{J. Chen},$^{	15					}$
{A. Ciszewski},$^{	13					}$
{R. Cruz-Torres},$^{	1					}$
{B. Dabrowska},$^{2}$
{D. Dabrowski},$^{2,18}$
{A. Dmitriev},$^{	2					}$
{A. Dryablov},$^{	2					}$
{P. Dulov},$^{	2					}$
{D. Egorov},$^{	2					}$
{A. Fediunin},$^{	2					}$
{I. Filippov},$^{	2					}$
{K. Filippov},$^{	10					}$
{D. Finogeev},$^{	16	,	10			}$
{I. Gabdrakhmanov},$^{	2					}$
{A. Galavanov},$^{	2	,	10			}$
{I. Gasparic},$^{	17					}$
{O. Gavrischuk},$^{	2					}$
{K. Gertsenberger},$^{	2					}$
{A. Gillibert},$^{	8					}$
{V. Golovatyuk},$^{	2					}$
{M. Golubeva},$^{	16					}$
{F. Guber},$^{	16	,	11			}$
{Yu. Ivanova},$^{	2					}$
{A. Ivashkin},$^{	16	,	11			}$
{A. Izvestnyy},$^{	16					}$
{S. Kakurin},$^{	2					}$
{V. Karjavin},$^{	2					}$
{N. Karpushkin},$^{	16					}$
{R. Kattabekov},$^{	2					}$
{V. Kekelidze},$^{	2					}$
{S. Khabarov},$^{	2					}$
{Yu. Kiryushin	},$^{	2					}$
{A. Kisiel},$^{	18					}$
{D. Klimanskiy,$^{	2					}$
{V. Kolesnikov},$^{	2					}$
{A. Kolozhvari},$^{	2					}$
{A. Konstantinov,$^{	2					}$
{Yu. Kopylov	},$^{	2					}$
{I. Korover},$^{	3					}$
{L. Kovachev},$^{	2, 20					}$
{A. Kovalenko},$^{	2					}$
{Yu. Kovalev},$^{	2					}$
{A. Kugler},$^{	19					}$
{S. Kuklin},$^{	2					}$
{E. Kulish},$^{	2					}$
{A. Kuznetsov},$^{	2					}$
{E. Ladygin},$^{	2					}$
{N. Lashmanov},$^{	2					}$
{E. Litvinenko},$^{	2					}$
{S. Lobastov},$^{	2					}$
{B. L\"oher},$^{	4					}$
{Y.-G. Ma},$^{	15					}$
{A. Makankin},$^{	2					}$
{A. Maksymchyuk},$^{	2					}$
{A. Malakhov},$^{	2					}$
{I. Mardor},$^{	3					}$
{S. Merts},$^{	2					}$
{A. Morozov},$^{	2					}$
{S. Morozov},$^{	16	,	10			}$
{G. Musulmanbekov},$^{	2					}$
{R. Nagdasev},$^{	2					}$
{D. Nikitin},$^{	2					}$
{V. Palchik},$^{	2					}$
{D. Peresunko},$^{	14					}$
{ M. Peryt},$^{	2					}$
{O. Petukhov},$^{	16					}$
{Yu. Petukhov},$^{	2					}$
{S. Piyadin},$^{	2					}$
{V. Plotnikov},$^{	2					}$
{G. Pokatashkin},$^{	2					}$
{Yu. Potrebenikov},$^{	2					}$
{O. Rogachevsky},$^{	2					}$
{V. Rogov},$^{	2					}$
{K.~Ros\l{}on},$^{	2	,	18			}$
{D. Rossi},$^{	4					}$
{I. Rufanov},$^{	2					}$
{P. Rukoyatkin},$^{	2					}$
{M. Rumyantsev},$^{	2					}$
{D. Sakulin},$^{	2					}$
{V. Samsonov},$^{	10					}$
{H. Scheit},$^{	4					}$
{A. Schmidt},$^{	1					}$
{S. Sedykh},$^{	2					}$
{I. Selyuzhenkov},$^{	10					}$
{P. Senger},$^{	10					}$
{S. Sergeev},$^{	2					}$
{A. Shchipunov},$^{	2					}$
{A. Sheremeteva},$^{	2					}$
{M. Shitenkov},$^{	2					}$
{V. Shumikhin},$^{	10					}$
{A. Shutov},$^{	2					}$
{V. Shutov},$^{	2					}$
{H. Simon},$^{	6					}$
{I. Slepnev},$^{	2					}$
{V. Slepnev},$^{	2					}$
{I. Slepov},$^{	2					}$
{A. Sorin},$^{	2					}$
{V. Sosnovtsev},$^{	10					}$
{V. Spaskov},$^{	2					}$
{T. Starecki},$^{	18					}$
{A. Stavinskiy},$^{	9					}$
{E. Streletskaya},$^{	2					}$
{O. Streltsova},$^{	2					}$
{M. Strikhanov},$^{	10					}$
{N. Sukhov},$^{	2					}$
{D. Suvarieva},$^{	2					}$
{J. Tanaka},$^{	4					}$
{A. Taranenko},$^{	10					}$
{N. Tarasov},$^{	2					}$
{O. Tarasov},$^{	2					}$
{V. Tarasov},$^{	9					}$
{A. Terletsky},$^{	2					}$
{O. Teryaev},$^{	2					}$
{V. Tcholakov},$^{	20					}$
{V. Tikhomirov},$^{	2					}$
{A. Timoshenko},$^{	2					}$
{N. Topilin},$^{	2					}$
{B. Topko},$^{	2					}$
{H. T\"ornqvist},$^{	4					}$
{I. Tyapkin},$^{	2					}$
{V. Vasendina},$^{	2					}$
{A. Vishnevsky},$^{	2					}$
{N. Voytishin},$^{	2					}$
{V. Wagner},$^{	4					}$
{O. Warmusz},$^{	13					}$
{I. Yaron},$^{	3					}$
{V. Yurevich},$^{	2					}$
{N. Zamiatin},$^{	2					}$
{Song Zhang},$^{	15					}$
{E. Zherebtsova},$^{	16					}$
{V. Zhezher},$^{	2					}$
{N. Zhigareva},$^{	9					}$
{A. Zinchenko},$^{	2					}$
{E. Zubarev},$^{	2					}$
{M. Zuev},$^{	2					}$

{\noindent$^{1}$} Massachusetts Institute of Technology, Cambridge, Massachusetts 02139, USA.
{\noindent$^{2}$} Joint Institute for Nuclear Research, Dubna 141980, Russia.
{\noindent$^{3}$} School of Physics and Astronomy, Tel Aviv University, Tel Aviv 69978, Israel.
{\noindent$^{4}$} Institut f\"ur Kernphysik, Technische Universit\"at Darmstadt, 64289 Darmstadt, Germany.
{\noindent$^{5}$} Dubna State University, Dubna 141980, Russia.
{\noindent$^{6}$} GSI Helmholtzzentrum f\"ur Schwerionenforschung GmbH, Planckstr. 1, 64291 Darmstadt, Germany.
{\noindent$^{7}$} Helmholtz Forschungsakademie Hessen f\"ur FAIR, Max-von-Laue-Str. 12, 60438 Frankfurt, Germany.
{\noindent$^{8}$} IRFU, CEA, Universit\'{e} Paris-Saclay, F-91191 Gif-sur-Yvette, France.
{\noindent$^{9}$} Institute for Theoretical and Experimental Physics (ITEP), Moscow, Russia.
{\noindent$^{10}$} National Research Nuclear University MEPhI, Moscow, Russia.
{\noindent$^{11}$} Moscow Institute of Physics and Technology (MIPT), Moscow, Russia.
{\noindent$^{12}$} Texas A\&M University-Commerce, Commerce, Texas 75429, USA.
{\noindent$^{13}$} University of Wroclaw, Wroclaw, Poland.
{\noindent$^{14}$} Kurchatov Institute, Moscow.
{\noindent$^{15}$} Key Laboratory of Nuclear Physics and Ion-Beam Application (MOE), Institute of Modern Physics, Fudan University, Shanghai, China.
{\noindent$^{16}$} Institute for Nuclear Research of the RAS (INR RAS), Moscow, Russia.
{\noindent$^{17}$} Rudjer Boskovic Institute, Zagreb, Croatia.
{\noindent$^{18}$} Warsaw University of Technology, Warsaw, Poland.
{\noindent$^{19}$} Nuclear Physics Institute, CAS, \v{R}e\v{z}, Czech Republic.
{\noindent$^{20}$} Plovdiv University ``Paisii Hilendarski'', Plovdiv, Bulgaria.
{\noindent$^{21}$} Bauman Moscow State Technical University.

\clearpage



\begin{center}
 {\noindent \large \bf Methods \\}
\end{center}

\bigbreak 
{\noindent \bf Ion beam.}
The primary beam ions were produced in a KRION source and accelerated in the Nuclotron~\cite{Kekelidze:2013cua},
delivered quasi-continuously in pulses for 
$2$ seconds followed by $8$ second pauses between spills.
Each pulse delivered $2.5\times10^5$~ions on average.

The beam contained a mixture of Carbon-12, Nitrogen-14, and Oxygen-16 ions with fractions on average of 68\%, 18\%, and 14\% respectively.
The $^{12}$C ions have a beam momentum of $3.98$~GeV/c/u at the center of the LH$_2$ target.
They are focused on the target with a beam diameter of about $4$~cm, See Extended Data Fig.~\ref{fig:lh2_vertex}c.

The beam ions are identified on an event-by-event basis using their energy loss in the BC detectors (BC1, BC2 upstream the target), which is proportional to their nuclear charge squared $Z^2$.
The selection of the incoming nuclear species is shown in Extended Data Fig.~\ref{fig:incPID}.
Pile-up events are rejected by checking the multiplicity of the BC2 time signal.

\bigbreak 
{\noindent \bf The detectors upstream the target.}
Prior to hitting the target the beam was monitored by the two thin scintillator-based beam counters (BC1, BC2) 
and two multi-wire proportional chambers (MWPCs) mentioned above. 
The MWPCs determined the incident beam ion trajectory for each event.
Besides using the energy deposition in the BCs for beam ion identification, the BC closer to the target was readout by a fast MCP-PMT 
used to define the event start time $t_0$. 
Beam halo interactions were suppressed using a dedicated BC veto counter (V-BC),
consisting of a scintillator with a $5$~cm diameter hole in its center. 

\bigbreak 
{\noindent \bf Liquid-hydrogen target.}
The target~\cite{BMN:Target} was cryogenically cooled and the hydrogen was recondensated using liquid helium. 
The liquid hydrogen was held at $20$ Kelvin and 1.1 atmospheres in a 30~cm long, 6~cm diameter, aluminized Mylar cylindrical container. 
The container entrance and exit windows were made out of $110$ micron thick Mylar. 
The target constitutes a 14\% interaction length for $^{12}$C.
A sketch of the target cell is shown in Extended Data Fig.~\ref{fig:lh2_vertex}.

\bigbreak 
{\noindent \bf Two-arm spectrometer (TAS).}
A two-arm spectrometer was placed downstream of the target and was used to detect the 
two protons from the $(p,2p)$ reaction that emerge between $24^\circ$ and $37^\circ$. 
The vertical acceptance of each arm is $\pm 7^\circ$.
These laboratory scattering angles correspond to $\sim 90^\circ$ ($75^{\circ}$ to $101^{\circ}$) QE scattering in the two-proton center-of-mass (c.m.) frame. 
Each spectrometer arm consisted of scintillator trigger counters (TC), 
gas electron multiplier (GEM) stations, and multi-gap resistive plate chamber (RPC) walls.

Proton tracks are formed using their hit locations in the GEM and RPC walls.
The vertex resolution along the beam-line direction is $1.8$~cm ($1\sigma$) and was measured using a triple lead-foil target as detailed in the Online Supplementary Material.

The time difference between the RPC and $t_0$ signals define the proton time of flight (TOF). The TOF, 
combined with the measured track length (accounting for the exact interaction vertex in the target), is used to determine its momentum.
Measurements of gamma rays from interactions with a single lead-foil target were used for 
absolute TOF calibration. An absolute TOF resolution of $175$ ps was extracted, which dominates the momentum resolution, see Online Supplementary Materials for details.

\bigbreak 
{\noindent \bf Data taking and quality.}
Signals from the TAS-TCs were combined with the BC and V-BC scintillator signals to form the main $^{12}$C$(p,2p)$ reaction trigger for the experiment.  Additional triggers were set up for monitoring and calibration purposes, see Online Supplementary Materials for details.  

The stability of the trigger was monitored online during the experiment as part of our data quality control. We collected and recorded about $20$ million triggers. As part of the beam monitoring and quality, the ratio between BC2/BC1 and BC4/BC3 was not smaller than $65\%$, and the rate on the V-BC is on average $24\%$ relative to BC2. The main $^{12}$C$(p,2p)$ reaction trigger had a rate of about $180$ Hz, as measured during live beam. Variations of BC pulse height over the measurement time was monitored and accounted for in the analysis. No significant run-to-run variations  were observed in any of the final observables.

\bigbreak
{\noindent \bf Reaction vertex and proton identification.}
The $z$-position (along the beamline) of the reaction vertex is reconstructed from two tracks in the TAS, while the $(x,y)$ position is obtained from the extrapolated MWPC track in front of the target (the latter provides a better transverse position resolution). 
Details about the algorithm and performance can be found in the Online Supplementary Materials.

The reconstructed vertex position along the beam-line and transverse to it with the liquid-hydrogen target inserted is shown in Extended Data Fig.~\ref{fig:lh2_vertex}. The structure of the target -- the LH$_2$ volume and other in-beam materials, such as the target walls, styrofoam cover, and various isolation foils -- is well reconstructed
The vertex quality is ensured by requiring that the minimum distance between the two tracks, which define the vertex, is smaller than $4$~cm.
In addition, we place a selection on the absolute $z$-vertex requiring it to be reconstructed within $\pm13$~cm from the center of the target.

Scattering from the target vessel that was not rejected by the veto counter is removed by a cut on the $(x,y)$-vertex direction. This removes a strong peak due to a styrofoam cover over the target (Extended Data Fig.~\ref{fig:lh2_vertex}c).

Having determined the tracks and the vertex, the momentum of each proton is calculated with respect to the incoming beam direction, using the TOF information between the target and the RPC.

In order to select $(p,2p)$ events from Quasi-Free Scattering (QFS), other particles which also create a track but originate from inelastic reactions, mostly pions, need to be rejected.
We apply several criteria (outlined in the next section), but the basic selection is a cut to the velocity of the two measured particles, shown in Online Supplementary Material Fig.~\ref{fig:vtx_mindist}b.
In the analysis, both particles detected in the TAS must pass a velocity condition $0.8<\beta<0.96$, removing fast and slow pions.

\bigbreak 
{\noindent \bf Fragment detection.}
Nuclear fragments following the $(p,2p)$ reaction are emitted at small angles 
with respect to the incident beam with momentum that is similar to the beam momentum.
To measure the fragment scattering angle, three silicon (Si) planes and two MWPCs are placed in the beam-line downstream the target.
Following the MWPCs the fragments enter a large acceptance $2.87$ T$\cdot$m dipole magnet, 
and are bent according to their momentum-to-charge ratio ($P/Z$), i.\,e. magnetic rigidity.
Two large-acceptance drift chambers (DCH) with 8 wire-planes each
are used to measure the fragment trajectory after the magnet.

The fragment momenta are determined from the measurement of their bending angle in the magnet.
Fragment identification (nuclear mass and charge) is done using their bend in the magnetic field and energy deposition in two scintillator BCs (3,4) placed between the target and the magnet entrance, see Fig.~\ref{fig:setup}b. 
The latter is proportional to the sum over all fragment charges squared, $Z_{\mathrm{eff}} \equiv \sqrt{\sum Z^2}$.

\bigbreak 
{\noindent \bf Fragment momentum and identification.}
We follow a simulation-based approach to derive $P/Z$ from a multi-dimensional fit (MDF) to the measured fragment trajectories before and after the magnet.
The particle trajectory is determined using the MWPC-Si track before the magnet and the DCH track after the magnet. Both tracks serve as input for the $P/Z$ determination.

The momentum resolution was determined using unreacted $^{12}$C beam ions (from empty-target runs) and found to equal 0.78~GeV/c (1.6\%) (Online Supplementary Material Fig.~\ref{fig:mdf_resol}). This resolution is consistent with the resolution expected from events obtained with simulation that accounts for the incoming beam energy spread. 
Using beam-trigger events (see Online Supplementary Material) we verified that the momentum reconstruction resolution is the same when the $^{12}$C ions go through a full liquid-hydrogen target.
The achieved momentum accuracy is evaluated from simulation to equal 0.2\%.

The fragment tracking efficiency, including the detection efficiency of the upstream MWPC-Si, downstream DCH detectors, and track reconstruction and selection algorithm equals $\sim 40\%$.
See Online Supplementary Materials for details on the tracking algorithms and its performance.

Figure~\ref{fig:setup}b illustrates an example of this fragment identification from the experimental data using $P/Z$ obtained by the MDF vs. total charge measured in the scintillators.

This work focuses only on fragments with nuclear charge of 4 or larger with a single track matched between the upstream and downstream tracks. 
Although the charge of the fragments is only measured as an integrated signal in BC3 and BC4 counters, the Boron isotopes can be selected unambiguously since no possible combination of fragments could otherwise mimic a signal amplitude proportional to $\sum Z^2 = 25$. 
In the case of $^{10}$Be, the only other fragment of interest here with $Z_{\mathrm{eff}} = 4$, contamination from within the resolution is excluded by using the additional $P/Z$ information.
$^{10}$Be is the only possible fragment with $P/Z\sim10$ GeV/c in that region and is well separated.

Besides requiring a good vertex and single global-track events, we employ $Z_{\mathrm{eff}}$ and $P/Z$ selection criteria to identify $^{11}$B, $^{10}$B, or $^{10}$Be. 
A two-dimensional charge selection, as for the incoming charge, was applied here for BC3 and BC4. 
For the selection in $P/Z$ vs. $Z_{\mathrm{eff}}$ also a two-dimensional cut was applied as indicated in Fig.~\ref{fig:setup}b with a $\sim2\sigma$ selection in $P/Z$.

\bigbreak
{\noindent \bf Single heavy fragment detection efficiencies.}
As discussed above, this work is limited to reactions with a single heavy ($Z\geq4$) fragment in the final state. 
The detection of such a fragment depends on the ability of the fragment to emerge from the liquid hydrogen target without re-interacting, 
and our ability to identify its charge in the two BCs downstream of the target, and reconstruct its tracks before and after the magnet.
 
We extract the efficiencies for the charge and track reconstruction using beam-only data (i.e. no target vessel in the beam-line).   
We assume that, within the quoted uncertainties below, there is no difference between the efficiencies for detecting $Z=6$ and $Z=5$ and $4$ fragments. 

In order to determine the efficiency for determining the fragment's charge in the BCs downstream the target, we first select
incident $^{12}$C ions based on their energy loss in the BC1 and BC2 counters (see Extended Data Fig~\ref{fig:incPID}). 
We then examine the fraction of those $^{12}$C ions also identified by their energy loss in BC3 and BC4 downstream the target.
This fraction defines a charge identification efficiency of $\epsilon_Z = (83\pm6)\%$, where the uncertainty
is obtained from examining different energy-deposition cuts between $2-3\sigma$ on the Gaussian distribution in BC3 and BC4.
The standard deviation in efficiency from this cut variation relative to the mean value defines the uncertainty.  
The fraction of such $Z_{\mathrm{in}}=Z_{\mathrm{out}}=6$ events with a single reconstructed track and $P/Z=8$~GeV/c is equal to $(39.5^{+1.7}_{-2.6})\%$, determined in a $\pm2.2\sigma$ range with $\pm0.45\sigma$ range to account for the uncertainty. 
In case of $^{10}$Be fragments the tracking efficiency is $(39.5^{+5.1}_{-7.8})\%$ due to larger systematic effects.
The larger asymmetry towards smaller efficiency arises from a possible background contribution in the reconstructed $P/Z$ that is taken into account.
More details are given below in ``Extracting QE ratios'' and in the Online Supplementary Material, in particular about a single-track identification and its efficiency.

\bigbreak
{\noindent \bf Single-proton knockout data-analysis.}
The basic selection for any analysis requires an incoming $^{12}$C, a good reaction vertex, and particles in the arms passing the velocity condition. 
These selections criteria define the inclusive $(p,2p)$ reaction channel, which is dominated by FSI and IE scattering.
The exclusive reaction channel requires the additional detection of a $^{11}$B fragment, with a single global-track condition and defines the one-proton QFS, that includes both QE and IE scattering.

We select a bound $^{11}$B where the $3/2^-$ ground-state is populated with the largest cross section.
However, we cannot distinguish bound excited states that de-excite via $\gamma$-ray emission that are also populated in our experiment.
Previous works ~\cite{Panin:2016div} found the contribution from such states to be small,
coming primarily from the $1/2^-$ and $3/2^-$ states that contribute $\sim$10\% each to the total cross section.
This contribution also correspond to $p$-shell knockout and does not impact the resulting momentum distribution significantly.

In order to identify $(p,2p)$ QE events and reject IE events, we look at the missing energy and the in-plane opening angle of the two particles measured in the arms.
An elliptical cut denoted by $2\,\sigma$ is applied in each direction (Fig.~\ref{fig:MF_Em_Pm}). 
The standard deviation was obtained from a Gaussian fit to $E_{\mathrm{miss}}$ ($\sigma=0.108$~GeV) and $\theta_{p1}+\theta_{p2}$ ($\sigma=1.8^{\circ}$).

The missing energy is defined as $E_{\mathrm{miss}} = m_p - e_{\mathrm{miss}}$, where $e_{\mathrm{miss}}$ is the energy component of $\bar{p}_{\mathrm{miss}}$ in the rest frame of the $^{12}$C nucleus.
The boost from the laboratory system into the rest frame is applied along the incoming-beam direction taking into account the reduced beam energy at the reaction vertex.
The selection region for QE events is defined in the exclusive channel with fragment selection, in a $2\sigma$ ellipse as indicated in Fig.~\ref{fig:MF_Em_Pm}.
The IE part is defined from the remaining events within the other ellipse.
The same criteria are applied in the inclusive channel.
Correlations with other kinematical variables are shown in Extended Data Fig.~\ref{fig:supp_emth}.

The $M^2_{\mathrm{miss}}$ spectrum in Extended Data Fig.~\ref{fig:supp_angPP}a shows the squared missing mass for the exclusive channel before and after applying the QE cut, clearly showing that we select background-free QE events with a missing mass that equals the proton mass. 
A lower boundary in the squared missing mass of $M^2_{\mathrm{miss}}>0.47$~GeV$^2/c^4$ is applied. 
Since the chosen selection criteria 
might influence other kinematical variables of $\bar{p}_{\mathrm{miss}}$ (Eq.~\ref{eq:Mmiss_MF_p2p}), we show the momentum distributions and angular correlations with less strict selection in the Extended Data (Figs.~\ref{fig:supp_angPP}, \ref{fig:supp_pm}) which do not show a different behavior and are also described well by the simulation.

\bigbreak
{\noindent \bf Single-proton knockout simulation.}
We compare the quasielastic $^{12}$C$(p,2p)^{11}$B data to a MonteCarlo simulation for the proton quasielastic scattering off a moving $^{12}$C.
In the calculation, the $^{12}$C system is treated as spectator plus initial proton, ${\bf p}_{^{12}\mathrm{C}} = {\bf p}_{^{11}\mathrm{B}} + {\bf p}_i$.
The proton's initial momentum distribution in $^{12}$C is sampled from a theoretical distribution.
Note that all kinematical quantities discussed here correspond to the carbon rest-frame.

The momentum distributions are calculated in the eikonal formalism for quasi-free scattering as described in Ref.~\cite{Aumann:2013hga}.  In this work we compare the data to the momentum-distribution calculated without absorption effects, i.\,e. without multiple-scattering. Here we also compare to the same calculation that includes absorption effects from the imaginary part of the potential explicitly, calculated in the optical limit of Glauber theory. See in Extended Data Fig.~\ref{fig:theo_abs_noabs}.

The distorted waves are calculated from  the real and imaginary part of the optical potential for the interaction between proton and nucleus. The single particle wave function of the removed proton is generated from a Woods-Saxon potential with radius given by $R=1.2\cdot A^{1/3}$~fm and diffuseness $a=0.65$~fm, while the depth of the potential was adjusted to reproduce the removal  energy, $S_p=15.96$~MeV, of a proton from the $p_{3/2}$-shell. For the $^{12}$C nucleus a density distribution from electron scattering was used as input, assuming that is has the same profile for the proton and neutron densities. The density is of the form $\rho_{^{12}\mathrm{C}} = (1 + \alpha\cdot(r/b)^2 )\cdot\exp\left\lbrace - r^2 / b^2\right\rbrace$, with $\alpha = 1.4 $ and $b$ chosen so as to reproduce the RMS radius of the $^{12}$C, $b= 2.47$~fm.

Although the fragment selection removes events from FSI and we do not need to account for their scattering into measured phase space, we look at the calculation with absorption since the survival probability is larger if the knockout happens at the nuclear surface. This effect might create a difference from no distortions. However, the momentum distributions with and without absorption look very similar, see Ext. Data Fig.~\ref{fig:theo_abs_noabs}, and do not seem to have a large impact on the reconstructed  initial momentum distribution in a light system such as $^{12}$C.

In terms of the kinematics, we raffle $\vert{\bf p}_i\vert$ from the total-momentum distribution and randomize its direction.
The proton's off-shell mass is 
\begin{align}
m^2_{\mathrm{off}} = m^2_{^{12}\mathrm{C}} + m^2_{^{11}\mathrm{B}} - 2 m_{^{12}\mathrm{C}}\cdot 
\sqrt{ m^2_{^{11}\mathrm{B}} + \mathbf{p}^2_i} .
\end{align}
The two-body scattering between the proton in $^{12}$C and the target proton is examined in their c.m. frame.
The elastic-scattering cross section is parameterized from free $pp$ differential cross section data.
Following the scattering process, the two protons and $^{11}$B four-momenta are boosted back into the laboratory frame.

The two-arm spectrometer was placed such that it covers the symmetric, large-momentum transfer, $90^{\circ}$ c.m. scattering region.
Given the large forward momentum, the detectors cover an angular acceptance of $\sim24^{\circ}<\theta<37^{\circ}$ 
in the laboratory system which corresponds to $\sim75^{\circ}<\theta_{\mathrm{c.m.}}<101^{\circ}$ in the c.m. frame.

In order to compare the simulated data to the experimental distributions, the simulation is treated and analyzed in the same way as the experimental data. Experimental acceptances are included. Resolution effects  are convoluted to proton and fragment momenta.
The proton time-of-flight resolution $\Delta\mathrm{ToF}/\mathrm{ToF}$ is $0.95\%$ at 2~GeV/c and the angular resolution 5~mrad, while the fragment momentum resolution is 1.5\% and the angular resolution 1.1~mrad in the $x$ and $y$ directions.
The angular resolution of the incoming beam is 1.1~mrad.
The beam-momentum uncertainty, examined as Gaussian profile, does not significantly impact rest-frame momentum distribution as long as the same nominal beam momentum is used for extracting physical quantities (or observables) from the experimental data and the simulated events. 
However, the momentum distributions are dominated by the width of the input p-shell momentum distribution.
When comparing, the simulation is normalized to the integral of the experimental distributions.
We find overall good agreement between experiment and Monte Carlo simulation showing that the reaction mechanism and QE events sample the proton's initial momentum distribution in $^{12}$C.
Additional data-simulation comparison is shown in Extended Data Fig.~\ref{fig:supp_pm}.

\bigbreak
{\noindent \bf Extracting QE $^{12}{\rm C}(p,2p{\rm X})/^{12}{\rm C}(p,2p)$ ratios for $^{11}{\rm B},$ $^{10}{\rm B},$ and $^{10}{\rm Be}$.}
To extract the fraction of $(p,2p)$ events with a detected heavy fragment we need to apply several corrections to the number of measured events which do not cancel in the ratio. The ratio of the exclusive cross section with a detected fragment to the inclusive cross section is given by:
\begin{equation}
\frac{^{12}{\rm C}(p,2p){\rm X}}{^{12}{\rm C}(p,2p)} = \frac{R} { \epsilon_{Z} \times \epsilon_{\mathrm{track}} \times att},  
\label{eq:FRAC}
\end{equation}
where
\begin{itemize}
  \item $R$ is the measured ratio based on the number of QE events for each sample. 
  We added a cut on low missing momentum, $p_{\mathrm{miss}}<250\;{\rm MeV/c}$, in addition to the missing energy and in-plane opening angle cuts to clean up the inclusive $(p,2p)$ sample, and focusing at the region of small missing momentum.
  \item $\epsilon_{Z}$ is the outgoing fragment charge efficiency. 
  We consider a value of $\epsilon_Z=(83\pm6)\%$, see discussion above.
  \item $\epsilon_{\mathrm{track}}$ is the outgoing fragment tracking efficiency with all the selection cuts applied in a $\pm 2.2\sigma$ $P/Z$ range. 
  We consider a value of $\epsilon_{\mathrm{track}}=(39.5^{+1.7}_{-2.6})\%$ for $^{11,10}{\rm B},$ and $\epsilon_{\mathrm{track}}=(39.5^{+5.1}_{-7.8})\%$ for $^{10}{\rm Be}$, see discussion above.  
    \item $att$ is the attenuation of the outgoing fragment due to secondary fragmentation in the target. 
  After the reaction, the flux of the fragment depends on the remaining distance the fragment needs to travel in the target.
  The attenuation is given by the reduction of this flux
   \begin{equation}
     att = {\rm exp}(-\rho \sigma_{\mathrm{tot}} z),
   \end{equation}
   where $\rho$ is the target density and $\sigma_{\mathrm{tot}}$ the total reaction cross section.
   We evaluate the attenuation factor by taking an average over the 30~cm target length, using $\sigma_{\mathrm{tot}}=220\pm10\;{\rm mb}$ (assumed to be the same for $^{10}{\rm B},^{10}{\rm Be}$ within uncertainty), such that
   $att=0.87\pm0.01$.
   Additional break-up reactions due to material in the beam-line downstream the target were estimated (and scaled) based on the total cross section on carbon.
   The contribution to the secondary reaction probability is comparably small, in particular reactions from $^{11}$B to $^{10}$B or $^{10}$Be are negligible.
\end{itemize}
The total reaction cross section $\sigma_{\mathrm{tot}}$ is calculated in eikonal reaction theory~\cite{Hussein:1990fr} using the $^{11}$B harmonic-oscillator like density distribution and the $NN$ cross section at 4~GeV/c/u as the input.
In a benchmark test it reproduces the measured cross section for $^{11}$B$+^{12}$C at kinetic energy of 950~MeV/u~\cite{Ozawa:2001hb} while the beam energy has only a very small impact.
We consider the $\sim5\%$ systematic overestimate of eikonal cross sections compared to measurements as uncertainty.

From Eq.~\ref{eq:FRAC} we see that there are four individual contributions to the uncertainty in the ratio of $^{12}{\rm C}(p,2p{\rm X})/^{12}{\rm C}(p,2p)$: statistics $\Delta R$, efficiencies ($\Delta \epsilon_Z$ and $\Delta \epsilon_{\mathrm{track}}$) and attenuation $(\Delta att)$. 
In addition we have a systematic uncertainty due to the event selection cuts. Each event cut was modified over a given $\sigma$ range and the resulting change in the relative yield was taken as the systematic uncertainty.
The 2D $E_{\mathrm{miss}}$-angle cuts were varied as $(2\pm1/2)\sigma$, where both these quantities are described by a Gaussian.
The cut in missing momentum was varied according to the missing momentum resolution like $p_{\mathrm{miss}}<250\pm 50\;{\rm MeV/c}$. 
These uncertainties are quoted as symmetric uncertainties since in the simulation we did not observe a significant asymmetry in the measured quantities.  
Besides that, we also consider a possible background contribution in the $P/Z$ determination as additional asymmetric systematic uncertainty.
It is determined for each charge selection separately with a fit in shape of a second order polynomial to the $P/Z$ distribution under quasielastic conditions.
Since the fits with and without background contribution result in very similar goodness we chose to adapt the possible background as $2\sigma$ uncertainty.
Combining these contributions we obtain the following fractions given with statistical (stat) and systematic (sys) uncertainties:
\begin{align*}
\frac{^{12}{\rm C}(p,2p)^{11}{\rm B}}{^{12}{\rm C}(p,2p)} &= (43.7\pm 2.4\,(\mathrm{stat})^{+4.9}_{-5.8}\,(\mathrm{sys})) \%, \\
\frac{^{12}{\rm C}(p,2p)^{10}{\rm B}}{^{12}{\rm C}(p,2p)} &= (7.8\pm 1.0\,(\mathrm{stat})^{+1.3}_{-1.4}\,(\mathrm{sys})) \%, \\
\frac{^{12}{\rm C}(p,2p)^{10}{\rm Be}}{^{12}{\rm C}(p,2p)} &= (0.9\pm 0.4\,(\mathrm{stat})^{+0.2}_{-0.3}\,(\mathrm{sys}))\% . 
\end{align*}

\bigbreak
{\noindent \bf Selecting high-momentum SRC events.} 
We study SRC events by focusing on $^{12}{\rm C}(p,2p)^{10}{\rm B}$ and $^{12}{\rm C}(p,2p)^{10}{\rm Be}$ events. 
We start with the two-proton detection imposing the vertex and $\beta$ cuts mentioned above. The first cut applied to select SRC breakup events is to look at high-missing momentum, $p_{\mathrm{miss}}> 350\;{\rm MeV/c}$. 

The remaining event selection cuts are chosen following a GCF simulation of the $^{12}{\rm C}(p,2p)$ scattering reaction off high missing-momentum SRC pairs.
After applying the high-missing momentum cut, we look at the in-plane opening angle between the protons for different cases: (a) inclusive $^{12}{\rm C}(p,2p)$ events, (b) GCF simulated SRC events, (c) exclusive $^{12}{\rm C}(p,2p)^{10}{\rm B}$ events, and (d) exclusive $^{12}{\rm C}(p,2p)^{10}{\rm Be}$ events. 
The GCF predicts relatively large opening angles 
that guides our selection of in-plane lab-frame opening angle larger than 63$^{\circ}$ (that also suppresses contributions from inelastic reactions that contribute mainly at low in-plane angles).   

Next we apply a missing-energy cut to further exclude inelastic and FSI contributions that appear at very large missing-energies. To this end we examine the correlation between the missing energy and missing momentum, after applying the in-plane opening angle cut, for the full range of the missing momentum (i.\,e., without the $p_{\mathrm{miss}}>350\;{\rm GeV/c}$ cut), see Extended Data Fig.~\ref{fig:Em_Pm_cut}. 
We chose to cut on $-110<E_{\mathrm{miss}}<240\;{\rm MeV}$. 

To improve the selection cuts we use the total energy and momentum conservation in reactions at which we identified a fragment ($^{10}${\rm B} or $^{10}${\rm Be}). 
We can write the exclusive missing-momentum in these reactions as
\begin{equation}
    \bar{p}_{\mathrm{miss,excl.}} = \bar {p}_{^{12}{\rm C}} + \bar{p}_{tg} - \bar{p}_1 - \bar{p}_2 -\bar{p}_{^{10}{\rm B}(\rm Be)}.
    \label{eq:Mm_exc}
\end{equation}
Neglecting the center-of-mass motion of the SRC pair, the missing-mass of this 4-vector should be equal to the nucleon mass $m^2_{\mathrm{miss,excl.}} \backsimeq m^2_N$. 
The distributions for $^{12}{\rm C}(p,2p)^{10}{\rm B}$ and $^{12}{\rm C}(p,2p)^{10}{\rm Be}$ events that pass the missing-momentum, in-plane opening angle, and missing-energy cuts are shown in Extended Data Fig.~\ref{fig:10B_Mm_exc} together with the GCF simulation. 
To avoid background events with very small values of the missing-mass we choose to cut on $M^2_{\mathrm{miss,excl.}}>0.42\;{\rm GeV^2/c^{4}}.$ 
After applying this cut we are left with 23 $^{12}{\rm C}(p,2p)^{10}{\rm B}$ and 2 $^{12}{\rm C}(p,2p)^{10}{\rm Be}$ events that pass all the SRC cuts. 

We note that if the measured SRC events were caused by FSI with a neutron in $^{11}$B, 
we would expect to also detect a similar number of $^{10}$Be fragments due to FSI with a proton in $^{11}$B.
At the high energies of our measurement these two FSI processes have almost the same rescattering cross sections~\cite{Alkhazov:1978et}. 
Our measurement of only $2$ $^{10}$Be events is consistent with the SRC $np$-dominance expectation and not with FSI.

In addition, while our selection cuts suppress QE scattering events off the tail of the mean-field momentum distribution they do not completely eliminate them. Therefore, some events could result from de-excitation of high-${p}_{\mathrm{miss}}$ $^{11}$B fragments. Using the de-excitation cross-sections of Ref.~\cite{Panin:2016div} and the measured number of $^{12}$C$(p,2p)^{11}$B events that pass our SRC selection cuts (except for the exclusive missing-mass cut), we estimate a maximal background of $4$ $^{10}$B and $2$ $^{10}$Be events due to knockout of mean-field protons and subsequent de-excitation.

\bigbreak
{\noindent \bf Characterizing the selected $^{12}{\rm C}(p,2p)^{10}{\rm B}$ events.} 
The majority of SRC events with a detected fragment comes with $^{10}{\rm B}$. 
In the Extended Data we present some kinematical distributions of these selected events together with the GCF simulation. 
Extended Data Fig.~\ref{fig:10B_Pm} shows the total $^{10}{\rm B}$ fragment and missing moments as well as their different components. 
Overall good agreement between the data and simulation is observed.

The pair c.m. momentum width of $\sigma_{\mathrm{c.m.}}=(156\pm27)$~MeV/c was obtained from the distribution in the transverse direction to the beam 
by $\chi^2$ comparison for several different c.m. width in the GCF simulation.
The result is consistent with electron scattering measurements~\cite{Cohen:2018gzh}.

Due to the high momenta of the nucleons in SRC pairs, it is beneficial to also analyze
the missing-momentum distribution in the relativistic light-cone frame where the 
longitudinal missing-momentum component is given by $\alpha = (E_{\mathrm{miss}} - p_{\mathrm{miss}}^z) / m_p$.
Similar to $p_{\mathrm{miss}}$, $\alpha$ is calculated in the $^{12}$C rest frame where $\hat{z}$ is boosted target-proton direction.
$\alpha = 1$ for scattering off standing nucleons.
$\alpha < 1$ ($> 1$) corresponds to interaction with nucleons that move along (against) the
beam direction and therefore decrease (increase) the c.m. energy of the reaction $\sqrt{s}$.
Extended Data Fig.~\ref{fig:10B_openAng_Pmn}a shows the $\alpha$ distribution for the measured SRC events.
We observe that $\alpha < 1$, as predicted by the GCF and expected given the strong $s$-dependence of the
large-angle elementary proton-proton elastic scattering cross-section.
For completeness, Extended Data Fig.~\ref{fig:10B_openAng_Pmn} also shows additional angular correlations between the nucleons in the pair and the $^{10}$B fragment, all well reproduced by the GCF.

\bigbreak
{\noindent \bf Estimating the number of SRC $^{12}{\rm C}(p,2p)^{10}{\rm B}$ and  $^{12}{\rm C}(p,2p)^{10}{\rm Be}$ events.}
As a consistency check we performed a simple estimate of the expected  number of exclusive SRC events based on the measured mean-field $^{12}{\rm C}(p,2p)^{11}{\rm B}$ event yield.
We assume SRCs account for $20\%$  of the wave function~\cite{egiyan06,wiringa14,Weiss:2016obx}, and that their contribution to the exclusive measurements is suppressed by a factor of $2$ as compared to the mean-field $^{12}{\rm C}(p,2p)^{11}{\rm B}$ due to the transparency of the recoil nucleon~\cite{hen12a,Duer:2018sjb,Dutta:2012ii}. Therefore, we expect a contribution of $11\%$ SRC and $89\%$ mean-field. 

The mean-field has contributions leading to bound states (i.\,e. $p$-shell knockout leading to $^{11}$B) and continuum states ($s$-shell knockout, non-SRC correlations, etc.) with relative fractions of $53$\% and $36$\% respectively ($53\%+36\%=89\%$)~\cite{Panin:2016div}. 
Therefore, given that we measured 453 $^{12}{\rm C}(p,2p)^{11}{\rm B}$ MF ($p$-shell knockout) events, we expect a total of $453 \cdot (11\%/53\%) = 94$ SRC events. 

We estimate the experimental loss due to acceptance of the longitudinal momentum (see Extended Data Fig.~\ref{fig:10B_Pm}a) as $50\%$, and another loss of $50\%$ due to the strong cuts applied to select SRC events. 
Thus, in total, we expect to detect about $94 \cdot 50\% \cdot 50\% = 24$ SRC events.

If the SRC pair removal results in $A-2$ fragments close to its ground-state, and assuming $np$-dominance 
(20 times more $np$ than $pp$ pairs) we expect a population of $90\%$ $^{10}{\rm B}$ and $10\%$ $^{10}{\rm Be}$. 
We also considered that for a $pp$ pair the knockout probability is twice larger than for $pn$. 
Using the estimation of 24 total SRC events will lead to 22 events for $^{10}{\rm B}$ (we measure 23) 
and 2 events for $^{10}{\rm Be}$ (we measure 2). These simple estimates show overall self-consistency in our data.

Last, as our selection cuts suppress, but do not eliminate events originating from the tail of the mean-field distribution, 
some events could result from de-excitation of high-$p_{\mathrm{miss}}$ $^{11}{\rm B}$ fragments. 
To evaluate that fraction, we consider $^{11}{\rm B}$ events that pass the SRC selection cuts 
(except for the exclusive missing mass cut). 
28 such events are observed of the total 453 MF $^{11}{\rm B}$ events (i.\,e. a fraction of $9\%$).
Reference~\cite{Panin:2016div} measured a neutron (proton) evaporation cross-section relative to the total continuum cross-section of $17\%$ ($7$\%). 
Using these fractions we expect a $^{10}{\rm B}$ $(^{10}{\rm Be})$ contribution from neutron (proton) evaporation based on the measured $^{11}{\rm B}$ events of $(28/53\%) \cdot 36\% \cdot 17\% = 3$ events ($(28/53\%) \cdot 36\% \cdot 7\% = 1$).
This is the maximum number that can be expected from this background, since for $^{10}{\rm B}$ and $^{10}{\rm Be}$ 
we apply an additional cut on the exclusive missing mass as explained above.

\bigbreak
{\noindent \bf GCF simulations.}
The GCF was derived and validated against many-body Quantum Monte Carlo calculations in Refs.~\cite{Weiss:2015mba,Weiss:2016obx,Cruz-Torres:2019fum}.  
Its implementation into an event generator that can be used for analysis of experimental data is detailed in Ref.~\cite{Pybus:2020itv},
and was successfully applied to the analysis of electron scattering SRC measurements in Refs.~\cite{Weiss:2018tbu,Duer:2018sxh,schmidt20,Pybus:2020itv}.

The adaptation of the GCF event generator from $(e,e'p)$ reactions to $(p,2p)$ reactions is simple and
mainly required replacing the electron mass with a proton mass when calculating the reaction kinematics
and phase-space factors and replacing the elementary electron-nucleon cross-section by the elastic proton-proton cross-section used in the mean-field simulation discussed above. 
We accounted for the experimental acceptance and detector resolution in the same way as described for the mean-field simulation discussed above.

The input parameters of the GCF calculation include: an $NN$ interaction model, for which we used the $AV18$ interactions;
consistent nuclear contact terms, that were taken from Ref.~\cite{Cruz-Torres:2019fum}; the width of the SRC pair c.m. momentum distribution,
which we set equal to $\sigma_{\mathrm{c.m.}}^{\mathrm{GCF}} = 150$~MeV/c~\cite{Cohen:2018gzh}; and an $A-2$-system excitation energy,
which we set to zero.

The uncertainty on the GCF calculation stems from uncertainties in the values of the nuclear contact terms (taken from Ref.~\cite{Cruz-Torres:2019fum}), $\sigma_{\mathrm{c.m.}}^{\mathrm{GCF}} = \pm 20$~MeV/c, and the $A-2$-system excitation energy. 
The latter was taken as equal to $2$ or $5$~MeV, with an abundance of $10\%$ each.

\clearpage\clearpage
\renewcommand{\figurename}{\bf Extended Data Fig.}
\renewcommand{\tablename}{\bf Extended Data Table}
\setcounter{figure}{0}
\setcounter{table}{0}

{
\begin{widetext}

\begin{center}
 {\noindent \large \bf Extended Data \\}
\end{center}
\bigbreak

\begin{figure}[h]
\centering  
	\includegraphics[width=\linewidth]{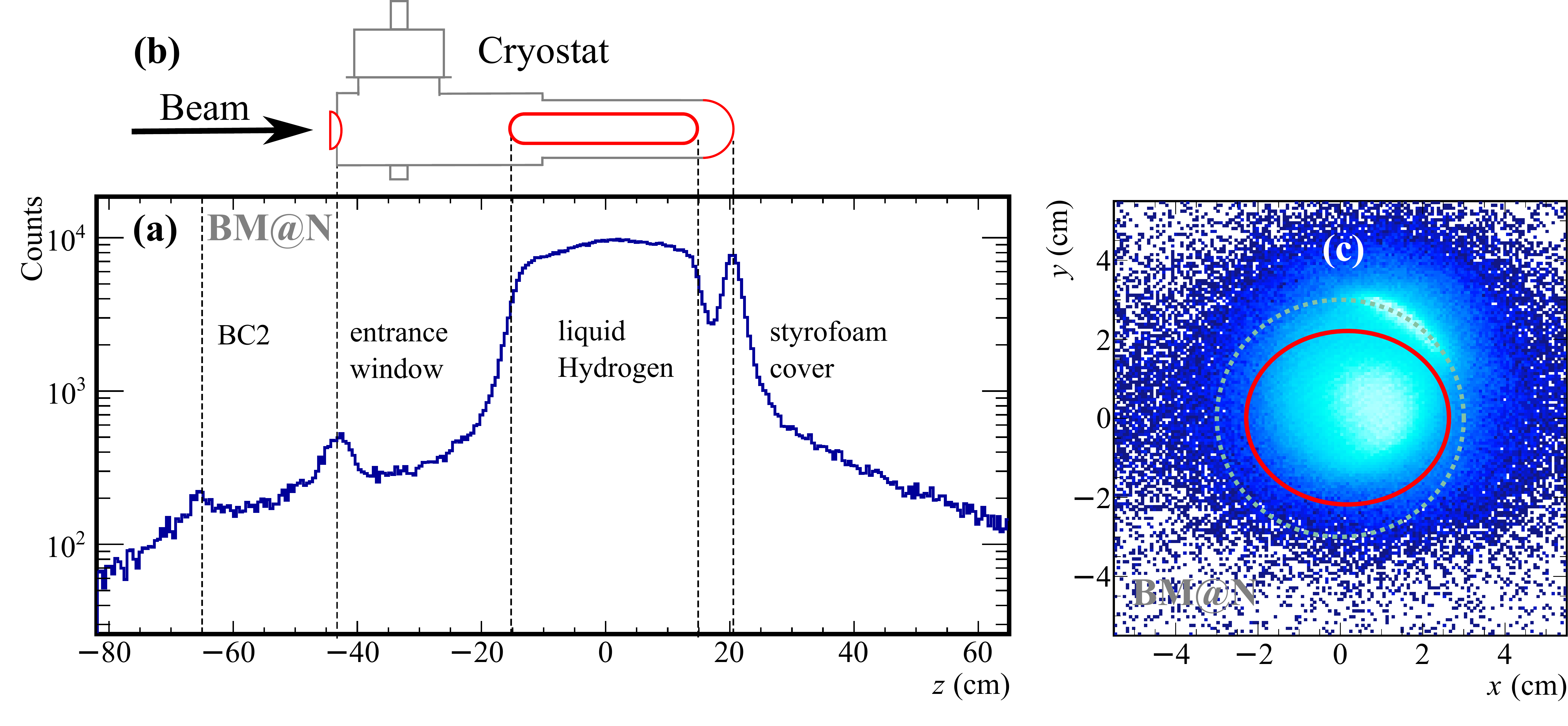}
	\caption{\bf $\vert$ Reaction vertex.
{\normalfont Reconstructed reaction vertex in the LH$_2$ target. 
The position along the beam line is shown in (a), scattering off in-beam material is also visible. 
For comparison, a sketch of the target device is shown in (b), scattering reactions are matched at the entrance window, the target vessel, styrofoam cover.
A selection in $z<\vert 13$~cm$\vert$ is applied to reject such reactions. 
The $xy$ position at the reaction vertex is shown in (c), measured with the MWPCs in front of the target.
The dashed line indicates the target cross section.
Scattering at the target vessel at around ($x=2$\,cm,$y=2$\,cm) can be seen which is removed by the selection as indicated by the red circle.}}
\label{fig:lh2_vertex}
\end{figure}

\begin{figure}[h]
\centering  
	\includegraphics[width=0.38\linewidth]{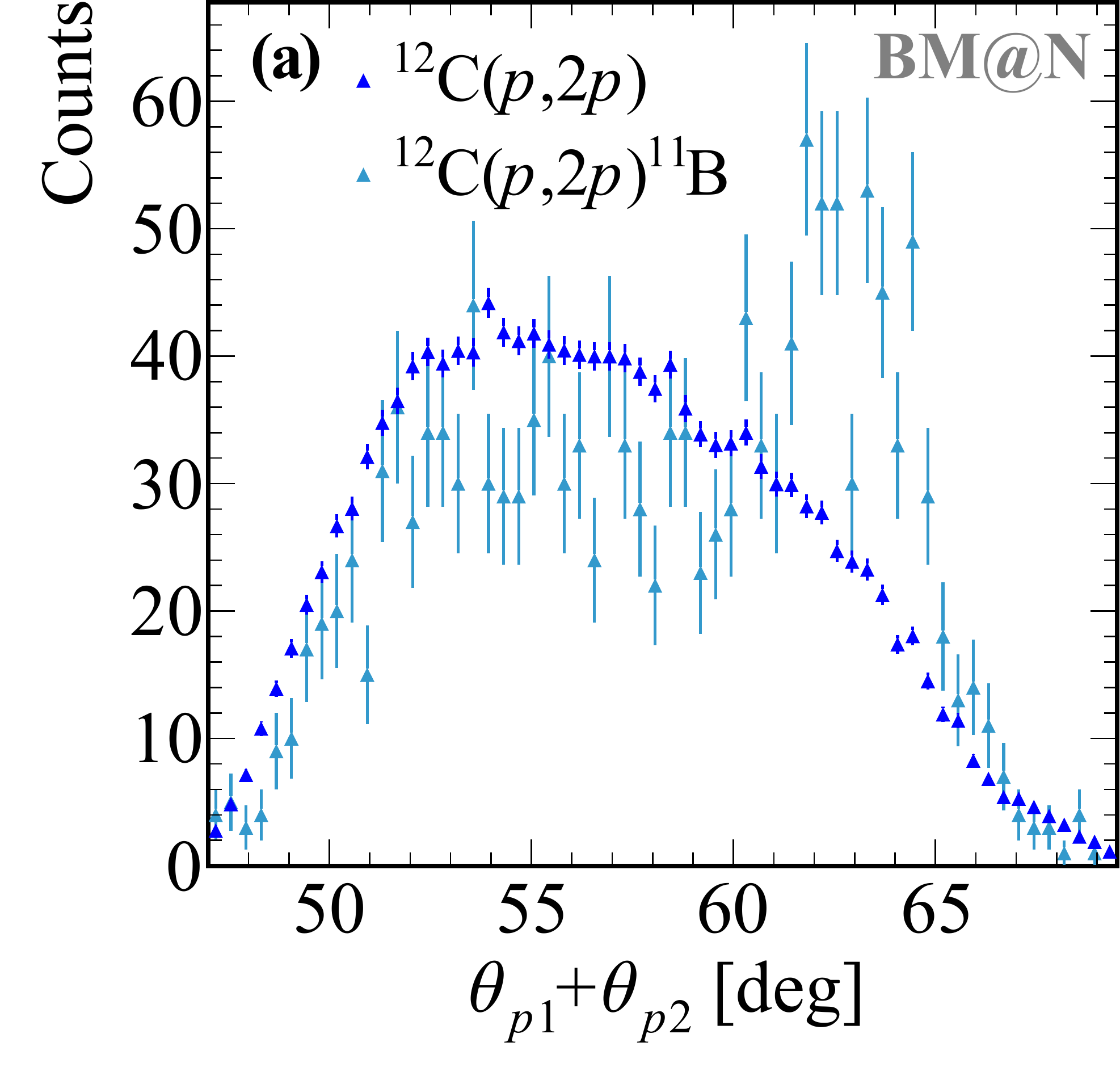}
	\hspace{0.2cm}
	\includegraphics[width=0.38\linewidth]{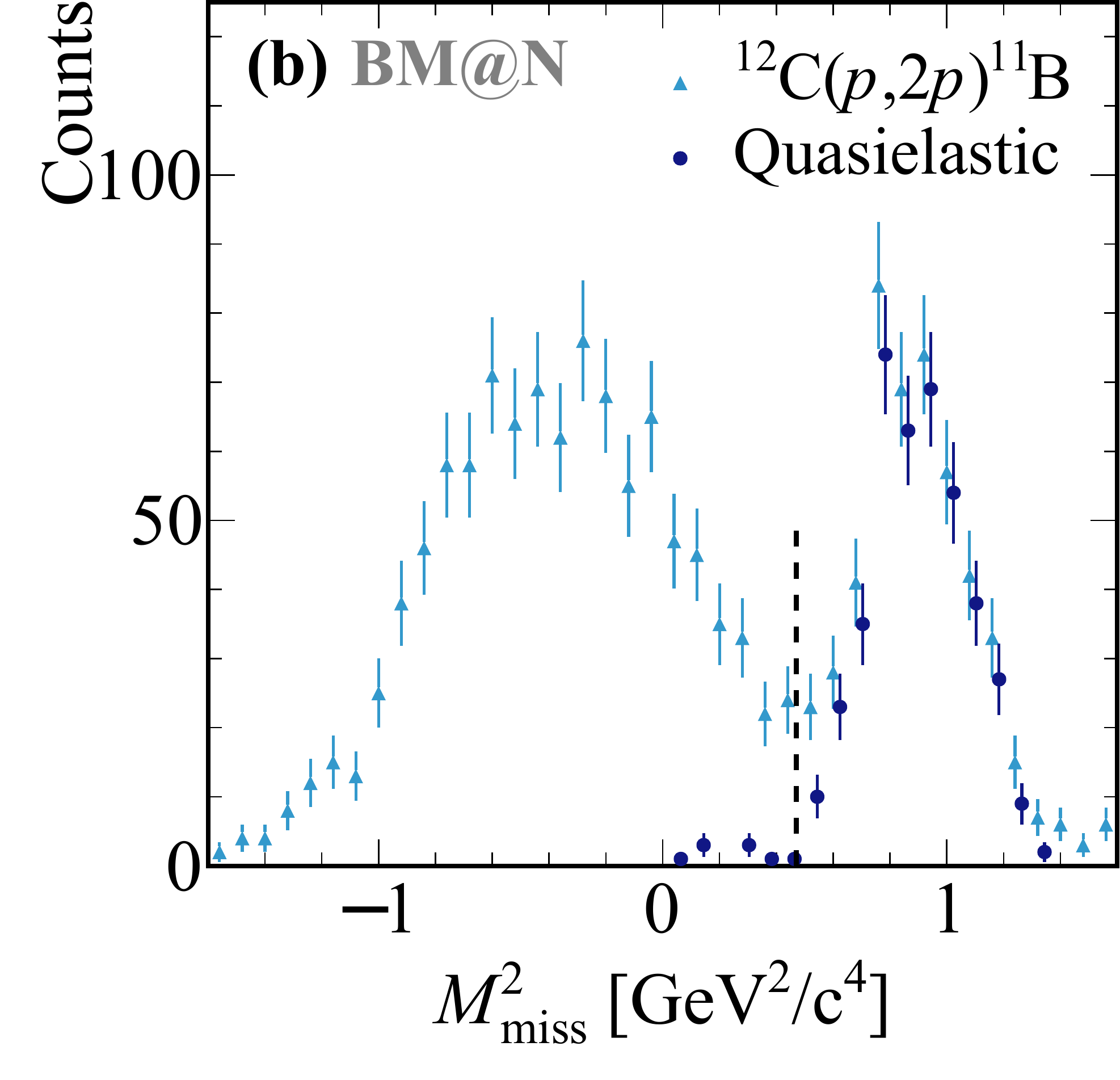}
	
	\vspace{0.5cm}
	
	\includegraphics[width=0.38\linewidth]{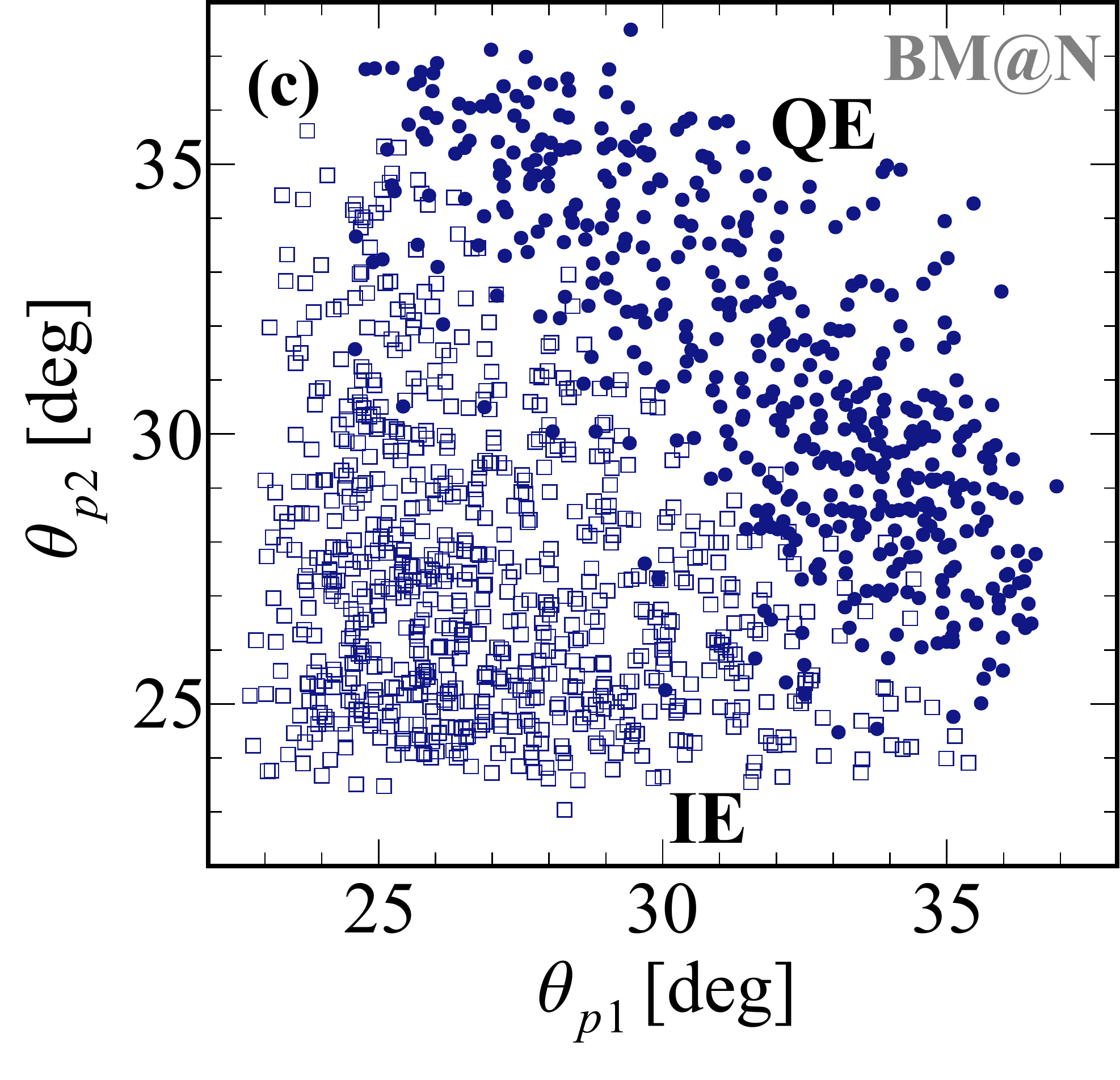}
	\hspace{0.2cm}
	\includegraphics[width=0.38\linewidth]{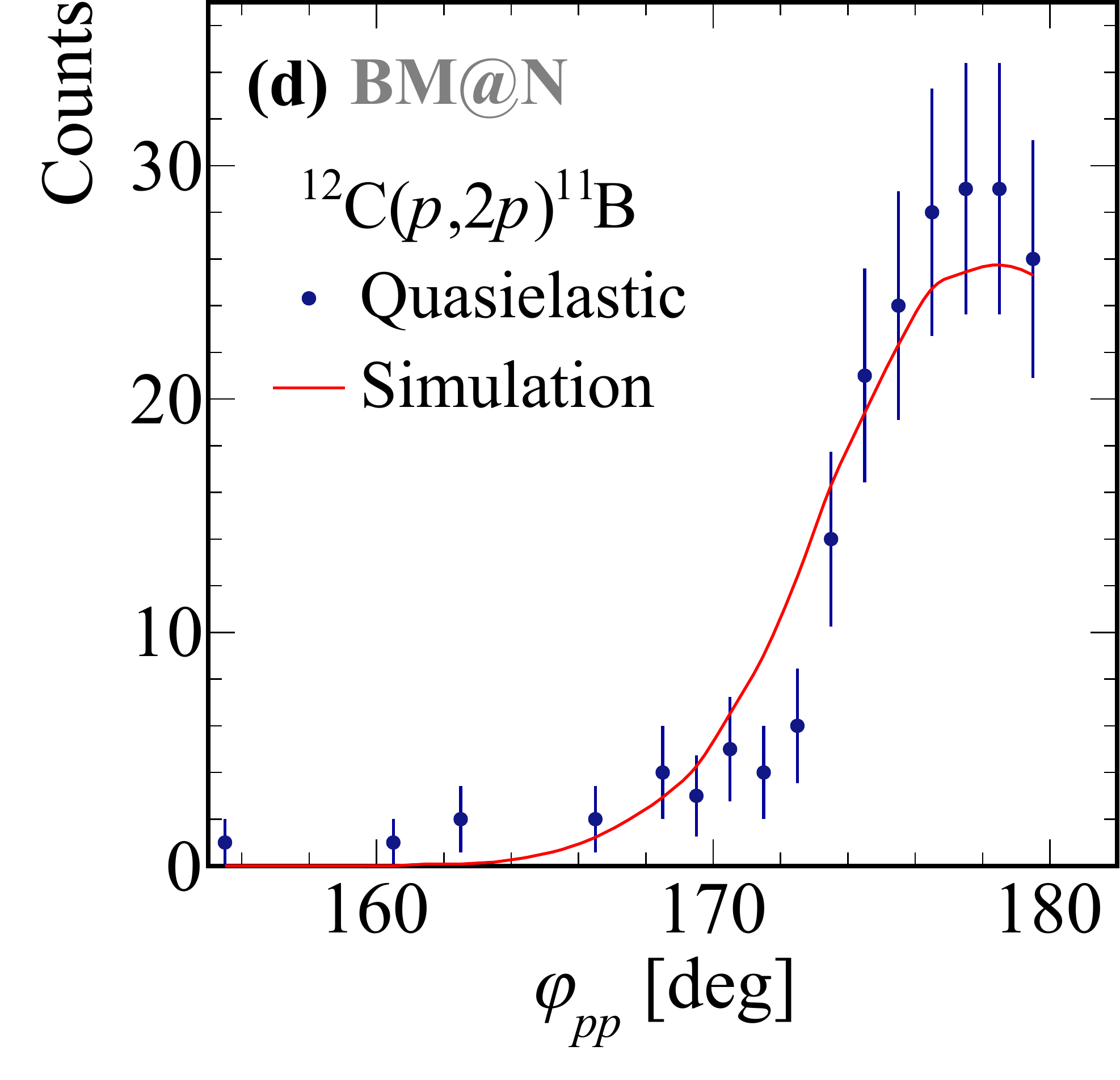}
	\caption{\bf $\vert$ Single-proton knockout signatures. 
	{\normalfont 
	(a) Projection for in-plane opening angle of Fig.~\ref{fig:MF_Em_Pm}, comparing the inclusive reaction $^{12}$C$(p,2p)$ and tagged events with $^{11}$B coincidence. The inclusive distribution is area normalized to the tagged one. The fragment selection clearly suppresses FSI, and the QE signal separates from IE. 
	(b) Proton missing mass for tagged $^{12}$C$(p,2p)^{11}$B events.
	After the QE selection in $E_{\mathrm{miss}}$ and in-plane opening angle, the distribution is shown in dark blue dots with artificial offset for better visibility.
	We apply an additional missing mass cut $M^2_{\mathrm{miss}}>0.47$~GeV$^2$/c$^4$, indicated by the dashed line. 
	(c) Angular correlation between the two $(p,2p)$ protons for quasielastic ($M^2_{\mathrm{miss}}>0.55$~GeV$^2$/c$^4$) and inelastic ($M^2_{\mathrm{miss}}<0.55$~GeV$^2$/c$^4$) reactions only selected by missing mass. 
	The QE events show a strong correlation with a polar opening angle of $\sim 63^{\circ}$. 
	(d) The off-plane opening angle peaks at 180$^{\circ}$ as expected, shown for $M^2_{\mathrm{miss}}>0.55$~GeV$^2$/c$^4$ . 
	The width of this distribution is narrower than that dictated by the TAS acceptance.
	Data error bars show the statistical uncertainties of the data at the 1$\sigma$ confidence level.
	}}
\label{fig:supp_angPP}
\end{figure}

\begin{figure}[h]
\centering  
	\includegraphics[width=\linewidth]{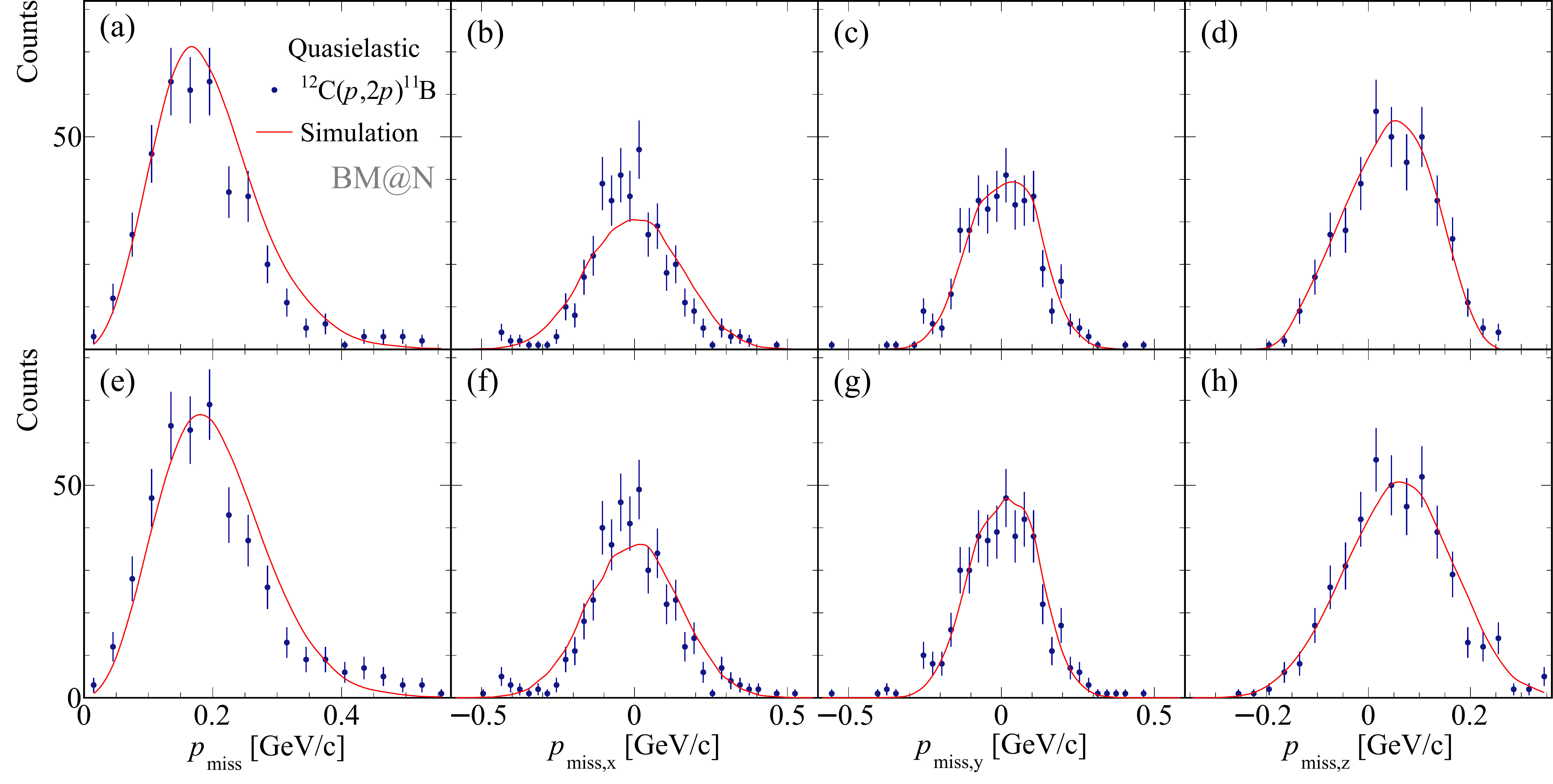}\\\qquad
	\includegraphics[width=\linewidth]{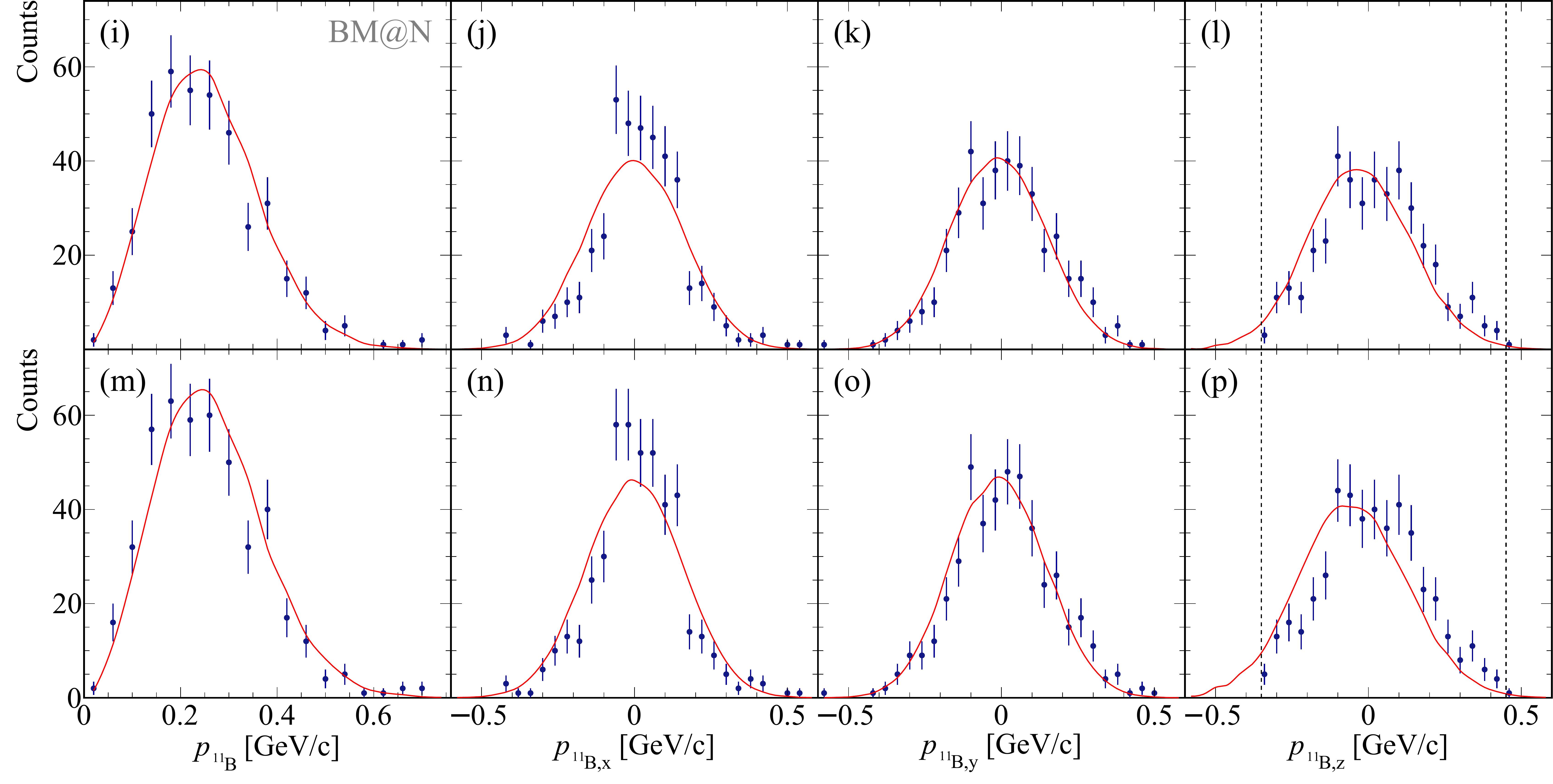}
	\caption{\bf $\vert$ Missing and fragment momentum. 
	{\normalfont Momentum components for quasielastic $^{12}$C$(p,2p)^{11}$B reactions compared to simulation. 
	The proton missing momentum is shown for (a)-(d), while (e)-(h) show the same distributions but with missing mass cut only (0.55~GeV$^2$/c$^4	<M^2_{\mathrm{miss}}<$1.40~GeV$^2$/c$^4$).
	Agreement with the simulation is found in both cases.
	The shift in $p_{\mathrm{miss},z}$ is associated with a strong $pp$ cross-section scaling with c.m. energy.
	For the same conditions the $^{11}$B fragment momentum components are shown in (i)-(l), and (m)-(p). 
	The dashed lines in $p_{^{11}\mathrm{B},z}$ indicate the momentum acceptance due to the fragment selection in $P/Z$.
	Data error bars show the statistical uncertainties of the data at the 1$\sigma$ confidence level.
	}}
\label{fig:supp_pm}
\end{figure}

\begin{figure}[h]
\centering  
	\includegraphics[width=\linewidth]{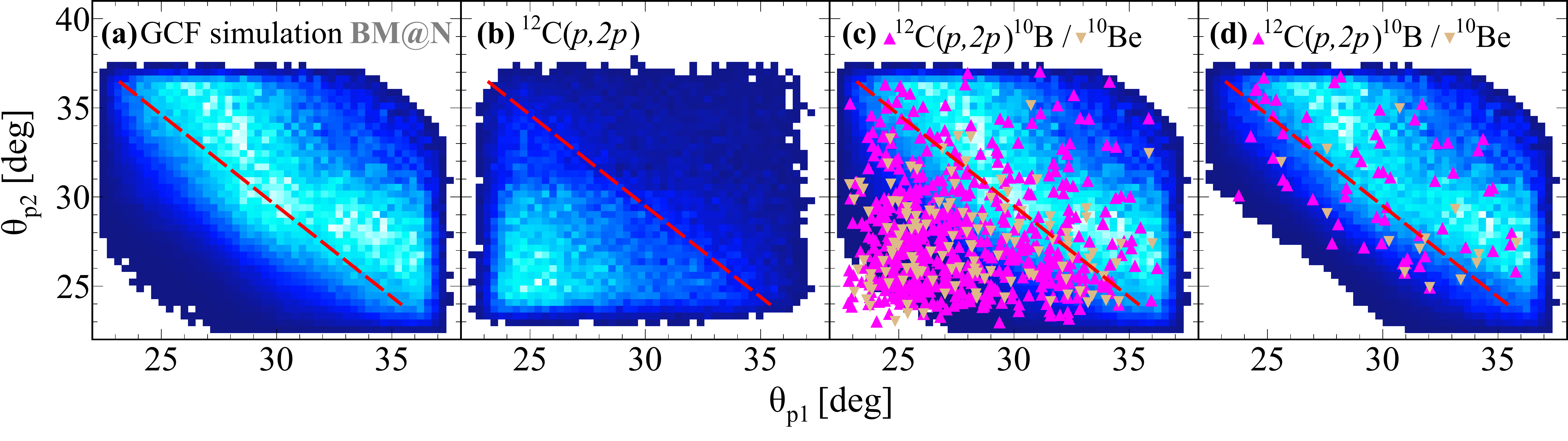}\\\qquad
	\includegraphics[width=\linewidth]{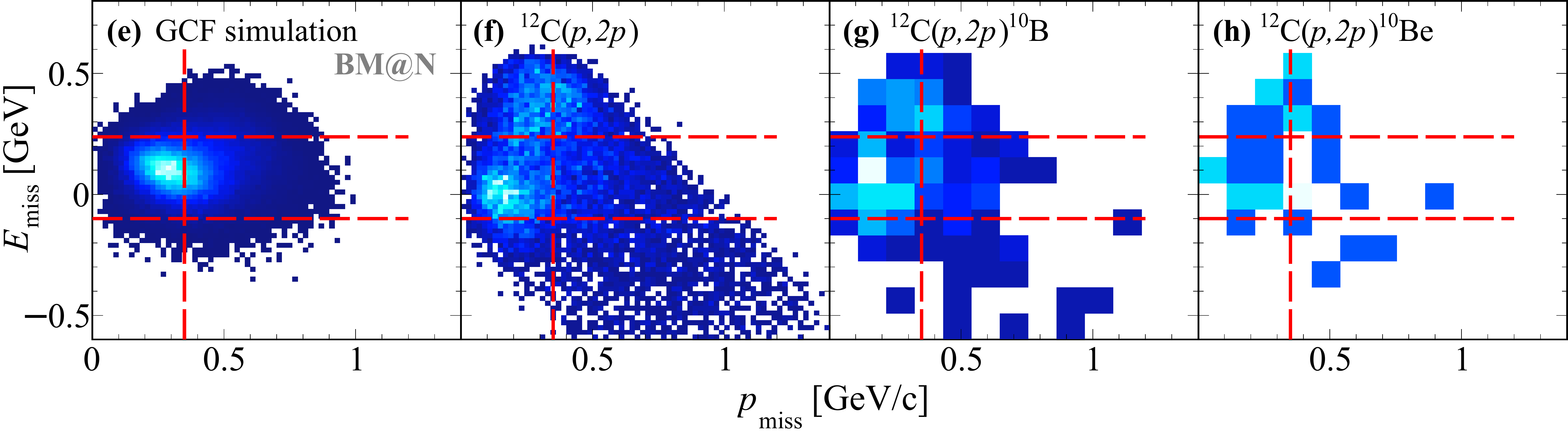}
	\caption{\bf $\vert$ SRC selection. 
	{\normalfont The proton-proton polar angular correlations are shown in (a)-(d) with $p_{\mathrm{miss}}>350$~MeV/c, the in-plane opening angle cut to be applied is indicated by the dashed line:
    (a) GCF simulation, (b) $^{12}{\rm C}(p,2p)$ data, (c) $^{12}{\rm C}(p,2p)^{10}{\rm B / Be}$ data on top of simulation, and (d) the same as (c) but with additional $E_{\mathrm{miss}}$ cut.
    The missing energy vs. missing momentum is shown in (e)-(h): for (e) GCF simulation, (f) $^{12}{\rm C}(p,2p)$, (g) $^{12}{\rm C}(p,2p)^{10}{\rm B}$, and (h) $^{12}{\rm C}(p,2p)^{10}{\rm Be}$ events that pass the in-plane opening angle cut.
	The selection cuts in $-110$~MeV$<E_{\mathrm{miss}}<240$~MeV and $p_{\mathrm{miss}}>350$~MeV/c are indicated by the dashed lines.}}
\label{fig:Em_Pm_cut}
\end{figure}

\begin{figure}[h]
\centering  
	\includegraphics[width=\linewidth]{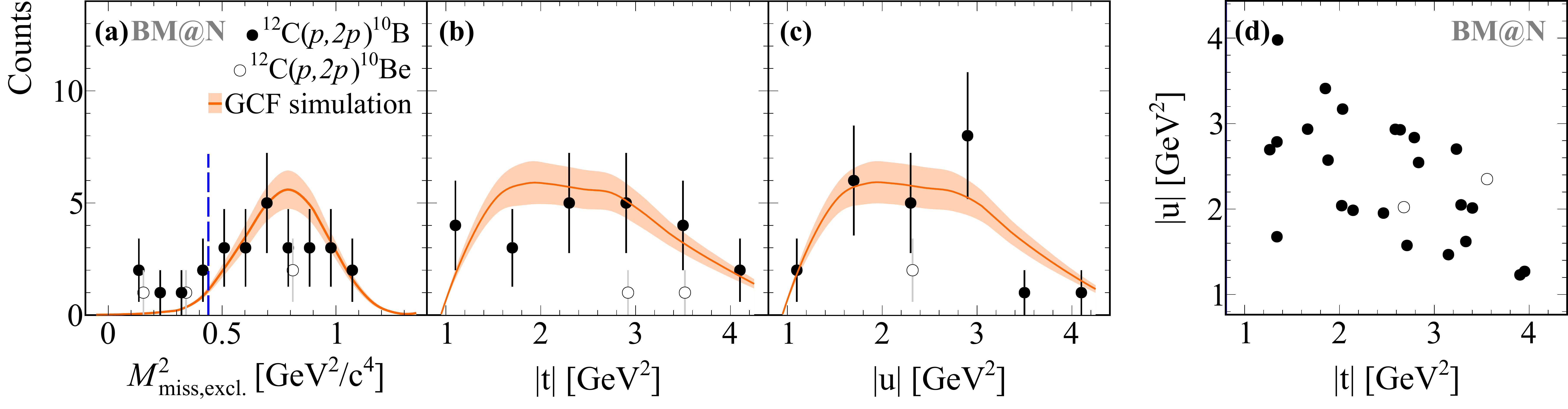}
	\caption{\bf $\vert$ SRC missing mass and momentum transfer.
	{\normalfont (a) The exclusive missing mass distributions for $^{12}{\rm C}(p,2p)^{10}{\rm B}$ events and $^{12}{\rm C}(p,2p)^{10}{\rm Be}$ events that pass the missing momentum, in-plane opening angle, and missing energy cuts together with the GCF simulation (orange). 
	The blue line represents the applied cut on the exclusive missing-mass $M^2_{\mathrm{miss,excl.}}>0.42\;{\rm GeV^2/c^4}$.
	(b) and (c) represent the Mandelstam variables for the same cases, $^{10}$B and $^{10}$Be, (d) shows the two-dimensional momentum-transfer plot for $^{10}$B.
			The width of the bands and the data error bars show the systematic uncertainties of the model and the statistical uncertainties of the data, respectively, each at the 1$\sigma$ confidence level.}}
\label{fig:10B_Mm_exc}
\end{figure}

\begin{figure}[h]
\centering  
		\includegraphics[width=\linewidth]{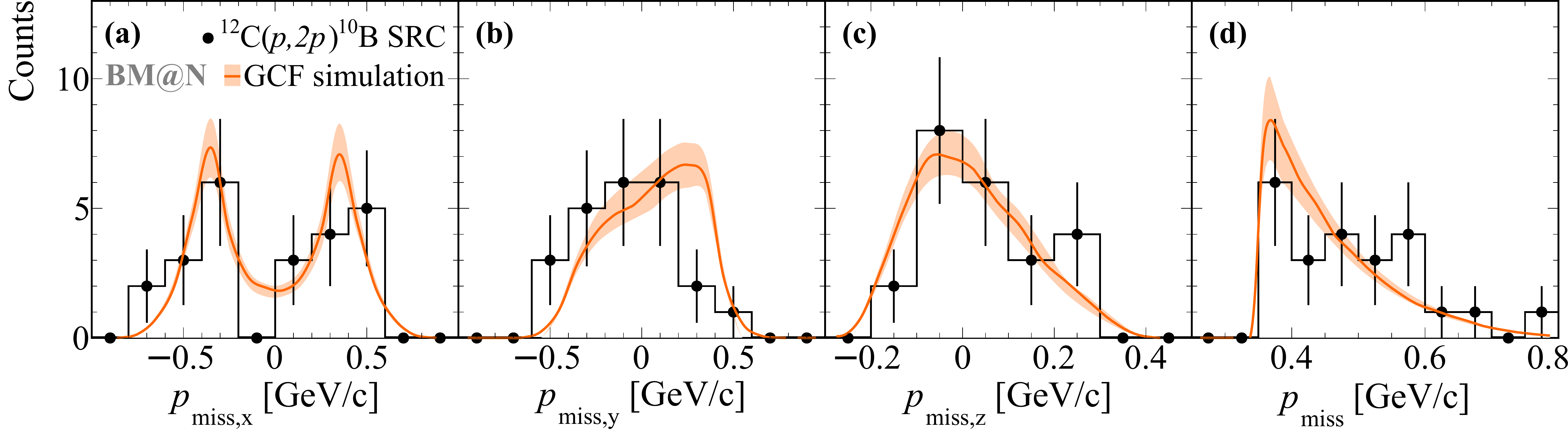}
		\vspace*{0.5cm}
		\includegraphics[width=\linewidth]{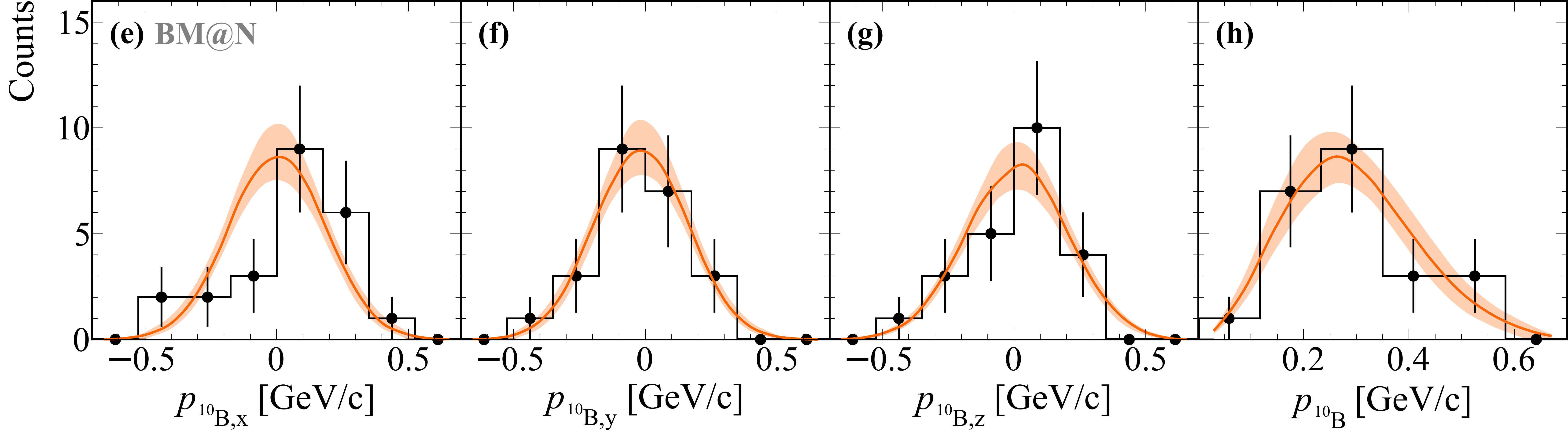}
	\caption{{\bf $\vert$ SRC missing and fragment momentum.}
	{\normalfont The missing momentum distributions (a)--(d) for the selected $^{12}{\rm C}(p,2p)^{10}{\rm B}$ SRC events (black) together with the GCF simulation (orange).
	Acceptance effects, especially in the transverse direction are well captured by the simulation. 
	The lower figures (e)--(h) show the fragment momentum distributions in the rest frame of the nucleus for the same selected $^{12}{\rm C}(p,2p)^{10}{\rm B}$ SRC events (black) together with the GCF simulation (orange).
		The width of the bands and the data error bars show the systematic uncertainties of the model and the statistical uncertainties of the data, respectively, each at the 1$\sigma$ confidence level.}}
\label{fig:10B_Pm}
\end{figure}

\begin{figure}[h]
\begin{center}
	\includegraphics[width=0.9\linewidth]{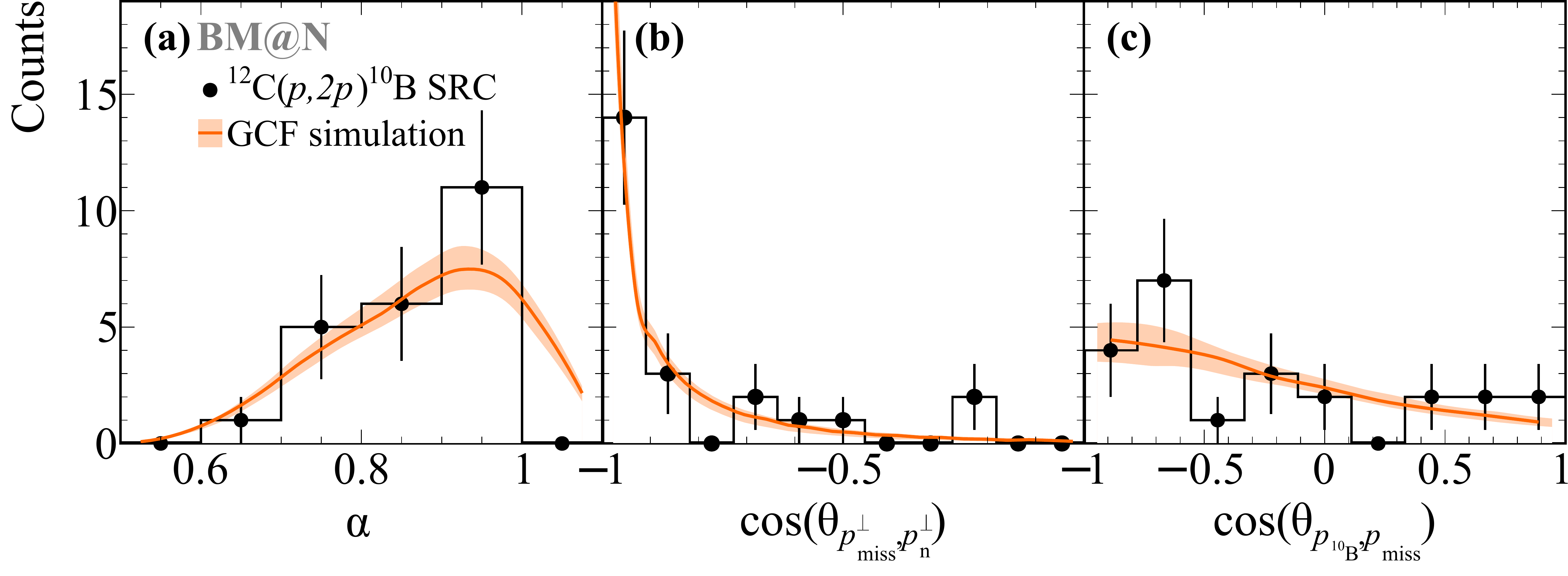}
	\caption{\bf $\vert$ SRC quantities.
{\normalfont Selected $^{12}{\rm C}(p,2p)^{10}{\rm B}$ SRC events (black) together with the GCF simulation (orange).
	(a) Light-cone momentum distribution $\alpha = (E_{\mathrm{miss}}-p^{z}_{\mathrm{miss}})/m_p$.
	(b) Cosine of the opening angle between the missing momentum and the neutron reconstructed momentum in the transverse direction.
	(c) Cosine of the angle between the $^{10}$B fragment and missing-momentum.
		The width of the bands and the data error bars show the systematic uncertainties of the model and the statistical uncertainties of the data, respectively, each at the 1$\sigma$ confidence level.
}}
\label{fig:10B_openAng_Pmn}
\end{center} 
\end{figure}

\begin{figure}[h]
\centering  
	\includegraphics[width=0.80\linewidth]{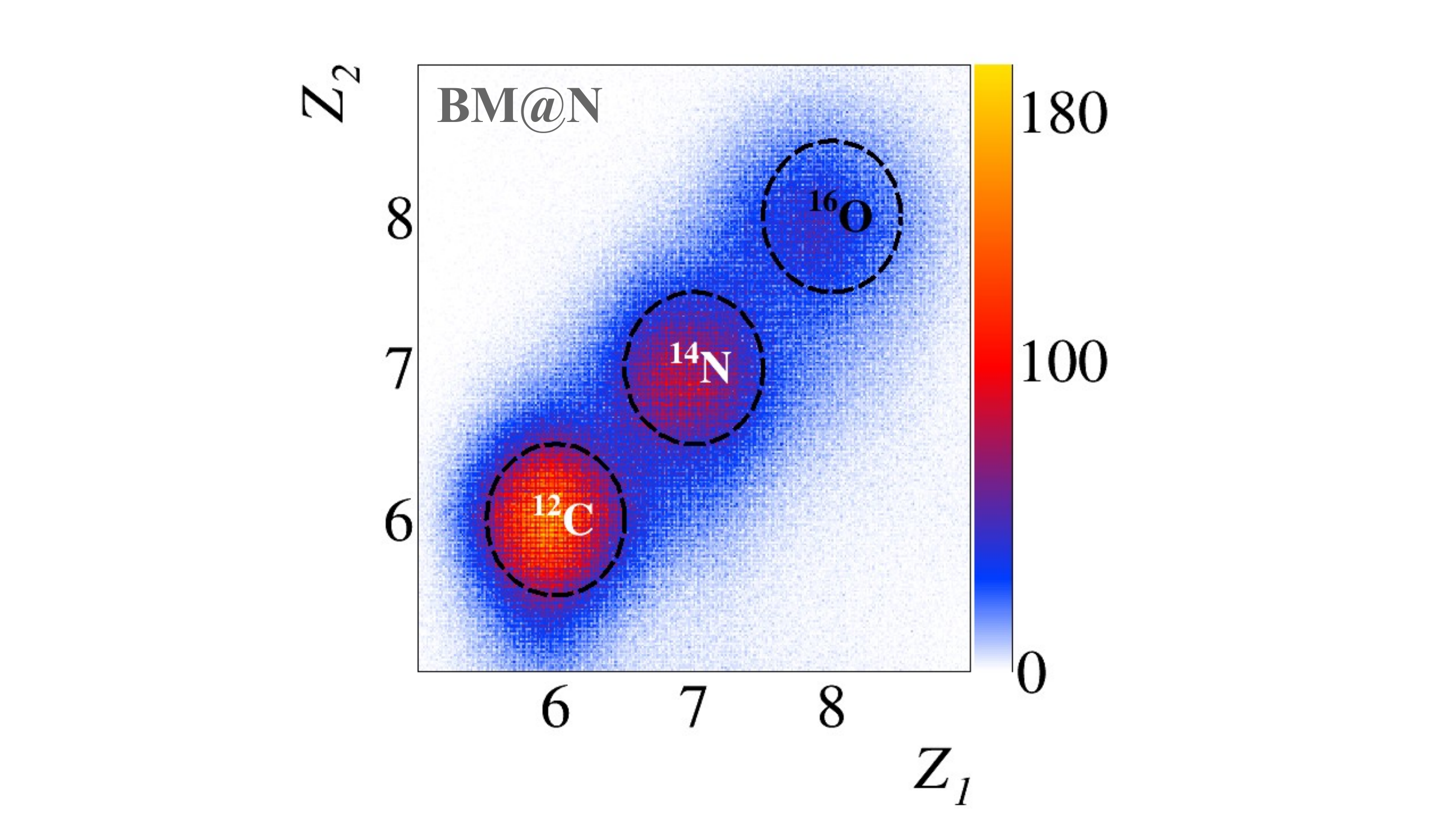}
	\caption{\bf $\vert$ Incoming beam ions.
{\normalfont Charge identification of incoming beam ions measured event-wise using the two BC counters in front of the target (BC1, BC2). 
Besides $^{12}$C, the $A/Z=2$ nuclei $^{14}$N and $^{16}$O are mixed in the beam with less intensity. 
}}
\label{fig:incPID}
\end{figure}

\begin{figure}[h]
\centering  
	\includegraphics[width=\linewidth]{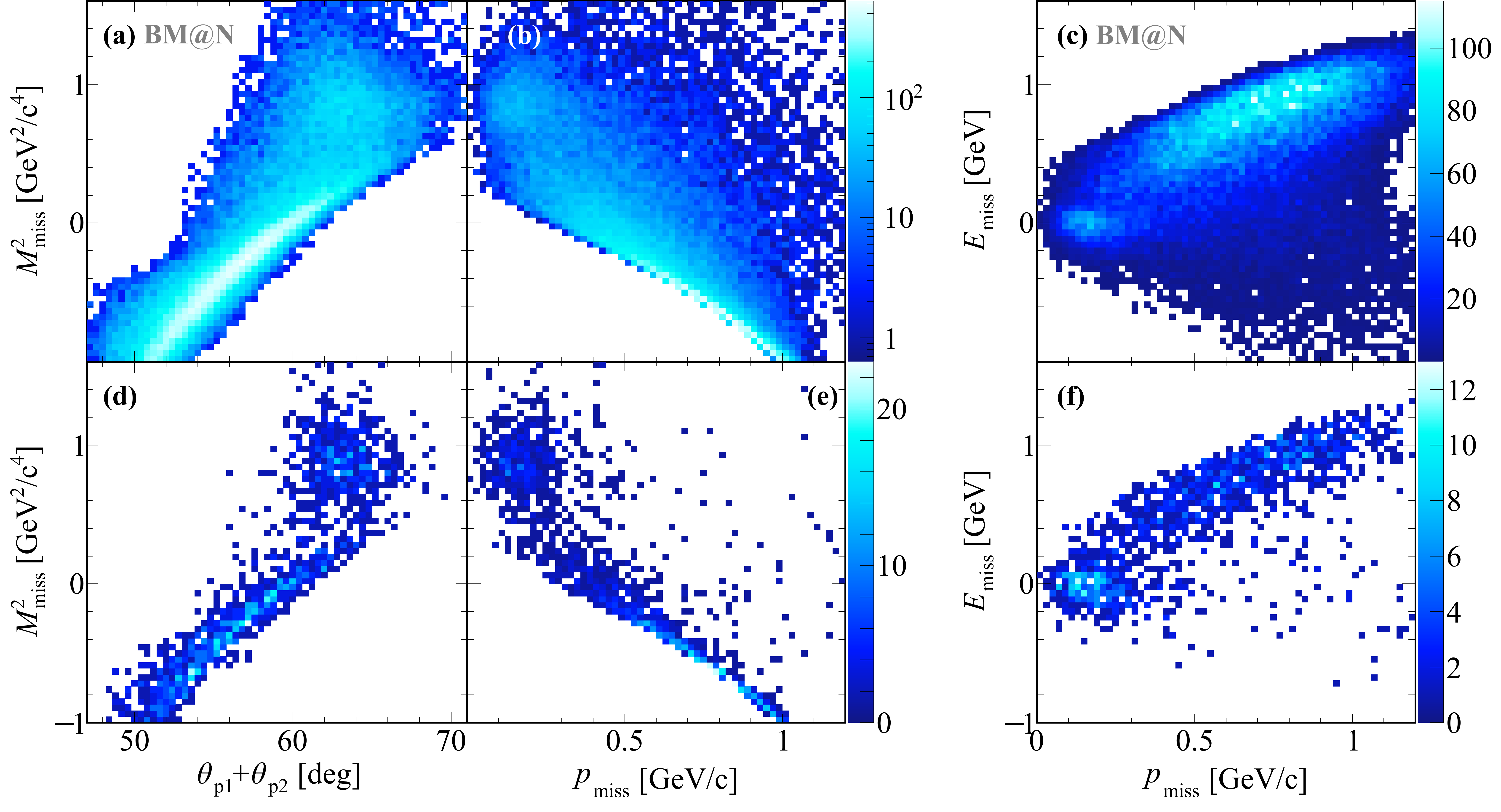}
	\caption{\bf $\vert$ Kinematical correlations in single-proton knockout. 
	{\normalfont Figures (a)-(c) show the inclusive $^{12}$C($p,2p$) channel, and (d)-(f) the exclusive channel, i.\,e. with tagging $^{11}$B. 
	In both cases, the quasielastic peak (QE) and inelastic (IE) events are visible, while ISI/FSI are reduced by the fragment tagging.
	Eventually, a selection in $E_{\mathrm{miss}}$ and in-plane opening angle was chosen to select QE events, see Fig.~\ref{fig:MF_Em_Pm}.
	The distributions are not corrected for fragment-identification efficiency.}}
\label{fig:supp_emth}
\end{figure}

\begin{figure}[h]
\centering  
	\includegraphics[width=0.45\linewidth]{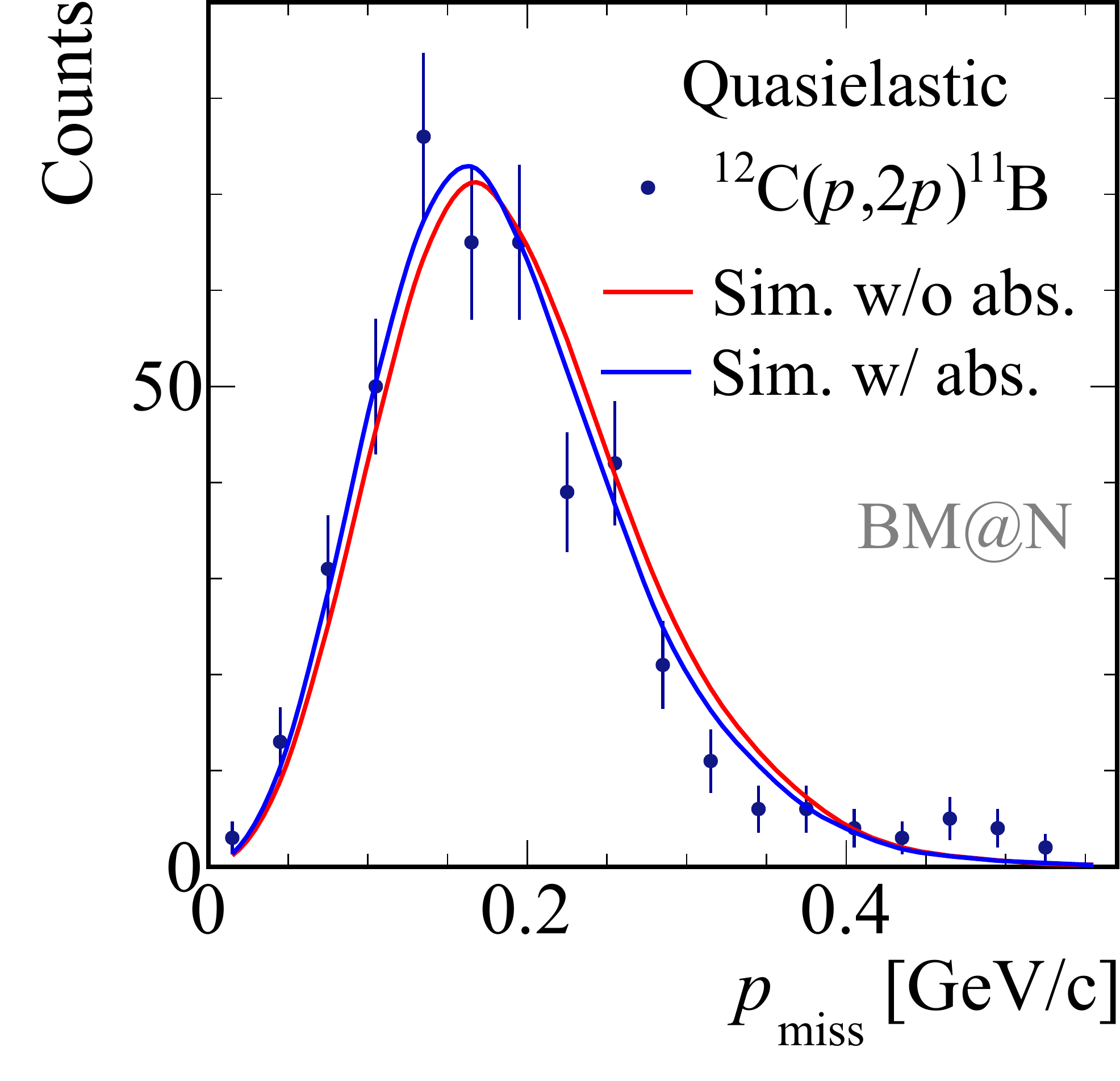}
	\caption{\bf $\vert$ Mean-field missing momentum calculations. 
	{\normalfont Missing-momentum distribution for quasielastic $^{12}$C$(p,2p)^{11}$B events, as in Fig.~\ref{fig:MF_Distributions} of the main text. 
	The data are compared with single-proton knockout simulation based on momentum distributions from an eikonal calculation with and without including absorption effects in the calculation and normalized to the same integral as the data.
	Both curves agree with the measured data and show only a small difference.
		Data error bars show the statistical uncertainties of the data at the 1$\sigma$ confidence level.}
	}
\label{fig:theo_abs_noabs}
\end{figure}

\clearpage
\clearpage
\renewcommand{\figurename}{\bf Supplementary Fig.}
\renewcommand{\tablename}{\bf Supplementary Table}
\setcounter{figure}{0}
\setcounter{table}{0}
\setcounter{equation}{0}

\begin{center}
 {\noindent \large \bf Supplementary Materials for: Unperturbed inverse kinematics nucleon knockout measurements with a 48~GeV/c Carbon beam\\}
\end{center}

\bigbreak
{\noindent \bf 1. BM@N detector configuration.}  
The BM@N experimental setup at JINR allows to perform fixed-target experiments with high-energy nuclear beams that are provided by the Nuclotron accelerator~\cite{Kapishin:2019wxa}.
Our experiment was designed such that in particular protons under large laboratory angles can be measured.
That dictated a dedicated upstream target position and modified setup as used for studies of baryonic matter, but using the same detectors~\cite{BMN:CDR}.
The setup comprises a variety of detection systems to measure positions, times, and energy losses to eventually obtain particle identification and determine their momenta.
We are using scintillator detectors, multi-wire proportional chambers, Silicon strip detectors, drift chambers, gas-electron multipliers, and resistive plate chambers as shown in Fig.~\ref{fig:setup} and described in the following.

{\bf Beam Counters (BC):} 
A set of scintillator counters, each based on a scintillator plate with an air light guide viewed by a PMT, was installed in the beam line. 
Two counters (BC1 and BC2) were located before the target: 
BC1 was located at the beam entrance to the experimental area.
It is a 15~cm in diameter and 3~mm thick scintillator read out by a XP2020 Hamamatsu PMT. 
BC2 was located right in front of the target and provided the start time $t_0$. 
This scintillator is of 4~cm x 6~cm x 0.091~cm size, and was tilted by 45$^{\circ}$ so that its effective area was around 4~cm x 4~cm. 
It was read out by a Photonis MCP-PMT PP03656. 
Two counters (BC3 and BC4), each read out by a XP2020 PMT, were located downstream the target to measure the total charge of the fragment particles in each event. 
BC3 was based on 10~cm x 10~cm x 0.29~cm scintillator, and the BC4 was 7~cm x 7~cm x 0.3~cm.
A veto-counter with the dimensions of 15~cm x 15~cm x 0.3~cm and a hole of 5~cm in diameter was located between BC2 and the target.
It was read out by an XP2020 PMT and was included in the reaction trigger to suppress the beam halo. 

{\bf Multi-wire proportional chambers (MWPC):}
We used two pairs of MWPC chambers, one before and one after the target for in-beam tracking~\cite{Khabarov:2019xjm}.
Each chamber has six planes 
$\left\lbrace\mathrm{X, U, V, X, U, V}\right\rbrace$. 
The X wires are aligned in $y$ direction, 
U and V planes are oriented $\pm60^{\circ}$ to X. 
The distance between wires within one plane is 2.5~mm, the distance between neighboring planes is 1~cm.
In total 2304 wires are read out. 
The active area of each chamber is 500~cm$^2$ (22~cm x 22~cm).
About 1~m separated the chambers in the first pair upstream the target and 1.5~m between the chambers in the second pair downstream the target. 
The polar angle acceptance of the chambers downstream the target is 1.46$^\circ$. 
The efficiency of the MWPC pair in front of the target for particles with the charge of 6 is (92.2$\pm$0.1)\%.
The efficiency of the MWPC pair after the target is (88.7$\pm$0.2)\% for ions with $Z = 6$, and (89.3$\pm$0.2)\% for ions with $Z = 5$.

{\bf Silicon trackers (Si):}
As additional tracking system, three Silicon planes~\cite{Kovalev_2017} were located after the target. 
In combination with the MWPCs after the target, an increased tracking efficiency is reached.
The first and second Si planes 
share the same housing. 
The first plane consists of four modules, the second plane has two modules, the third plane has eight modules. 
Each module has 640 $X$-strips (vertical in $y$-direction) and 640 $X^\prime$-strips (tilted 2.5$^{\circ}$ relative to $X$ strips). 
The first plane has smaller modules with 614 $X^\prime$ strips and 640 $X$ strips. 
The first two planes and the third plane are separated by 109 cm. 
The angular acceptance of the Si detector system is 1.58$^\circ$.
The design resolution of 1 mm for the $y$-coordinate and 50 $\mu$m for the $x$-coordinate was achieved in the experiment. 
The efficiency and acceptance of the Si tracking system, determined for reconstructed MWPC tracks before the target, is $(81.5\pm0.7)\%$ for outgoing $Z=6$ ions, and $(82.6\pm0.7)\%$ for $Z=5$ isotopes. 

Combined tracks
were reconstructed using information from the MWPC pair after the target and the Si detectors. 
The efficiency to find a Si track, and/or a track in the second pair of the MWPC, or a combined track is $(97.7\pm0.2)\%$ for $Z = 6$ ions, and $(97.9\pm0.3)\%$ for $Z=5$ isotopes evaluated for events with reconstructed tracks upstream the target. 
For the fragment tracking additional matching conditions are required with downstream DCH tracks, as explained below, which ensures additional good track selection.

{\bf Drift Chambers (DCH):}
Two large-area drift chambers, separated by 2~m, are located downstream the bending magnet. 
These detectors are used for tracking the charged fragments in the forward direction.
Together with the upstream-tracking information of MWPC and Si in front of the magnet, the bending angle and thus the magnetic rigidity of the ions is determined.
Each chamber consists of eight coordinate planes, twice $\left\lbrace\mathrm{X, Y, U, V}\right\rbrace$, where X wires are perpendicular to the $x$-axis, Y wires are at $90^{\circ}$ relative to X, and U and V are tilted by $+/-45^{\circ}$, respectively.
The distance between wires within one plane is 1~cm, in total 12,300 wires are read out. 
The spatial resolution, given as residual resolution, for one plane (X, Y, U, or V) is around 200~$\mu$m ($1\sigma$).
It is obtained by the difference between the measured hit and the position from the reconstructed track at that plane.
The efficiency of around 98\% (97\%) for each plane was estimated for the first (second) DCH based on the reconstructed matched track in the second (first) DCH.
A reconstructed track within one DCH chamber has at least 6 points.

{\bf Two-Arm Spectrometer (TAS):}
In order to detect light charged particles from the target, scattered to large laboratory angles, the symmetric two-arm detection system around the beamline was constructed for this experiment. 
Each arm, placed horizontally at $+/-29.5^{\circ}$ (center) with respect to the beamline, was configured by the following detectors along a 5~m flight length: scintillator -- scintillator -- GEM -- RPC.
Each arm holds one GEM (Gas-Electron Multiplier) station at a distance of 2.3~m from the target.
Each GEM station 
contained two GEM planes with the dimensions of 66~cm ($x$) x 40~cm ($y$) each, placed on top of each other (centered at $y=0$) to increase the overall sensitive area to 66~cm x 80~cm. 
The spatial resolution of the GEM hit is 300~$\mu$m.
Each RPC detector station, located at the end of the two arms at a distance of 5~m from the target, has a sensitive area of 1.1~m x 1.2~m. 
Each station consists of two gas boxes next to each other, each holds 5 multi-gap Resistive-Plate Chambers (RPCs) planes inside~\cite{Babkin:2016gjf}. 
Two neighboring planes within one box overlap by 5~cm in $y$ direction. 
Each plane has 30~cm long 1.2~cm wide horizontally aligned readout strips with a pitch of 1.25~cm. 
The measured $x$ position is obtained by the time difference measured between the ends of one strip.
The resolution is 0.6~cm.
Together with the position information from the GEM, tracks are reconstructed along the arms and the time-of-flight information is taken from the RPC system.
The clustering algorithm was applied to the neighboring strips fired in the same event.
In addition, each arm was equipped with two trigger counters (TC), scintillator planes close to the target. 
The X planes consisted of two scintillators with dimensions of 30~cm x 15~cm x 0.5~cm located vertically side by side and read out by a Hamamatsu 7724 PMT each. 
The distance between the target center and the X-counters was 42~cm.
Each Y plane was a single scintillator piece of 50~cm x 50~cm x 2~cm, read out by two ET9954KB PMTs. 
The distance between the target center and the Y planes was 170~cm.
Each arm covers a solid angle of 0.06~sr, limited by the RPC acceptance.

{\bf Data Acquisition System (DAQ) and Triggers:}
The DAQ performs readout of the front-end electronics of the BM@N detectors event-by-event based on the information of the trigger system~\cite{BMN:DAQ}. 
Timing information were read out from DCH and RPC (two-edge time stamp) and processed by Time to Digital Converters (TDC) based on HPTDC chip with typical accuracy of 20~ps for RPC and 60~ps for DCH. 
The amplitude information were read out from coordinate detector systems of Si and GEMs and processed by Amplitude to Digital Converters (ADC). 
The last 30~$\mu$s of waveforms were read back. 
The clock and time synchronization was performed using White Rabbit protocol. 
As mentioned in the main text, the reaction trigger was set up requesting an incoming ion in coincidence with signals in the left and right arm trigger scintillator-counters (TC).
Additional triggers are built from coincident signals in the various scintillator detectors, suited for either calibration purposes or data taking.
The trigger matrix is shown in Table~\ref{tab:triggers}, creating the so-called Beam trigger, and the physics triggers AndSRC and OrSRC.
The input signals are BC1, BC2, and no veto signal (!V-BC).
The coincidence condition AndXY requires signals in all TCs in the left and right arm, while OrXY takes the OR between the left and right arm of the spectrometer.
The physics data were taken requesting the AndSRC trigger at a rate of about 180~Hz as measured during a beam pulse duration, allowing a livetime close to 100\%.
\begin{table}[h]
\centering
\caption{$\vert$ Trigger matrix. {\normalfont Different coincidence triggers for collecting the data.}}
\begin{tabular}{llllll}
Trigger     & BC1 & BC2 & !V-BC & AndXY & OrXY \\
Beam        & x   & x   & x   &       &      \\
AndSRC      & x   & x   & x   & x     &      \\
OrSRC       & x   & x   & x   &       & x   
\end{tabular}
\label{tab:triggers}
\end{table}

\bigbreak
{\noindent \bf 2. Fragment momentum calculation} 
Trajectories of charged particles are bent in the large analyzer magnet according to their magnetic rigidity $B\rho$, i.\,e. momentum-over-charge ratio $B\rho=P/Q$ with charge $Q$. This allows to determine the fragment total momenta.

For this purpose, simulations of the fragments, propagating in the magnetic field, were carried out using the field map of the magnet. 
The corresponding materials of the beam-line detectors were also implemented in the simulation. 
The simulated fragments were chosen to have the maximum possible position, angular and momentum spread to cover the entire geometrical acceptance of the magnet and detectors. The output of the simulation is used afterwards as a training sample for the multidimensional fit (MDF) algorithm~\cite{RootMDF} in the form of n-tuples which hold positions and angles of the fragment trajectory upstream and downstream of the magnet: $( x_0,y_0,z_0, \alpha_x,\alpha_y)$   and  $(x_1,y_1,z_1 ,\beta_x,\beta_y) $ respectively.
Performing MDF over the training sample yields an analytical fit function $P/Z^{mdf}=f(x_0,y_0,z_0, \alpha_x,\alpha_y, x_1,y_1,z_1, \beta_x,\beta_y)$, which can be applied to the positions and angles measured in the experiment.

In a similar way, a second MDF function for $\alpha_x$ angle was derived as $\alpha_x^{mdf}=g(x_0,y_0,z_0, \alpha_y, x_1,y_1,z_1, \beta_x,\beta_y)$. 
This function is used for the track-matching condition $(\alpha_x^{mdf} - \alpha_x)$=min, which allows to determine whether the tracks in upstream and downstream detection systems belong to the same global track through the magnet.

Having determined the two functions, $\alpha_x^{mdf}$ and $P/Z^{mdf}$, experimental data for the reference trajectory of unreacted $^{12}$C is used to adjust the input variables' offsets, which reflect the alignment of the real detectors in the experimental setup with respect to the magnetic field. 
This is achieved by variation of the offsets in the experimental input variables simultaneously for $\alpha_x^{mdf}$ and $P/Z^{mdf}$ until the residual between $P/Z^{mdf}$ and its reference value is minimal. 
The reference  value is chosen to be the $P/Z$ of unreacted $^{12}$C at the exit of the liquid-hydrogen target.
Using this approach a total-momentum resolution of 0.78 GeV/c for $^{12}$C is achieved, as estimated with the empty target data, consistent with the resolution limits of the detection systems, see Fig.~\ref{fig:mdf_resol}. 
The same momentum resolution was obtained for unreacted $^{12}$C events, analyzed under the same conditions but with LH$_2$ target inserted. A width of $\sigma=0.78$~GeV/c was measured with a reduced beam momentum of 47.6~GeV/c due to energy loss in the target and additional straggling.
The achieved momentum accuracy is evaluated from simulation to be 0.2\%.  

Figure~\ref{fig:tx_matching} shows the performance of the second MDF function for $\alpha_x$. 
A global track is constructed when the reconstructed $\alpha_x^{mdf}$ falls within the 5$\sigma$ gate indicated.
In the analysis, only events with one heavy global track, which combines the up- and downstream detectors, are considered (if not stated differently).
To ensure that real detected single-track events are selected, a matching between the upstream and DCH angle in y direction is applied together with the above explained x-angle matching, also in a $5\sigma$ selection from their residual. 
Additionally, a single track in the DCH, the one reconstructed track from DCH1 and DCH2, is required.

\begin{figure}[]
\centering  
	\includegraphics[width=0.39\linewidth]{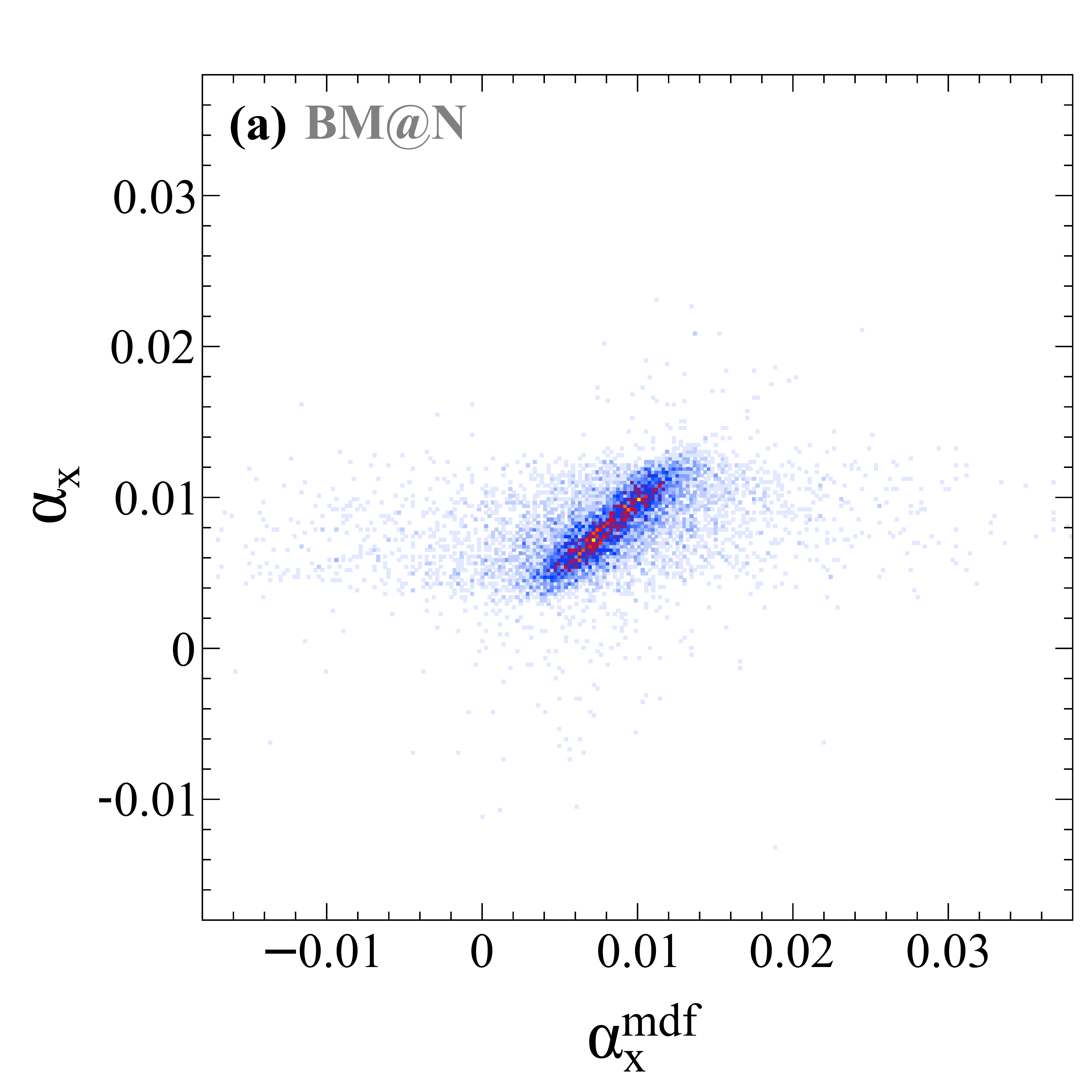}
	\includegraphics[width=0.4\linewidth]{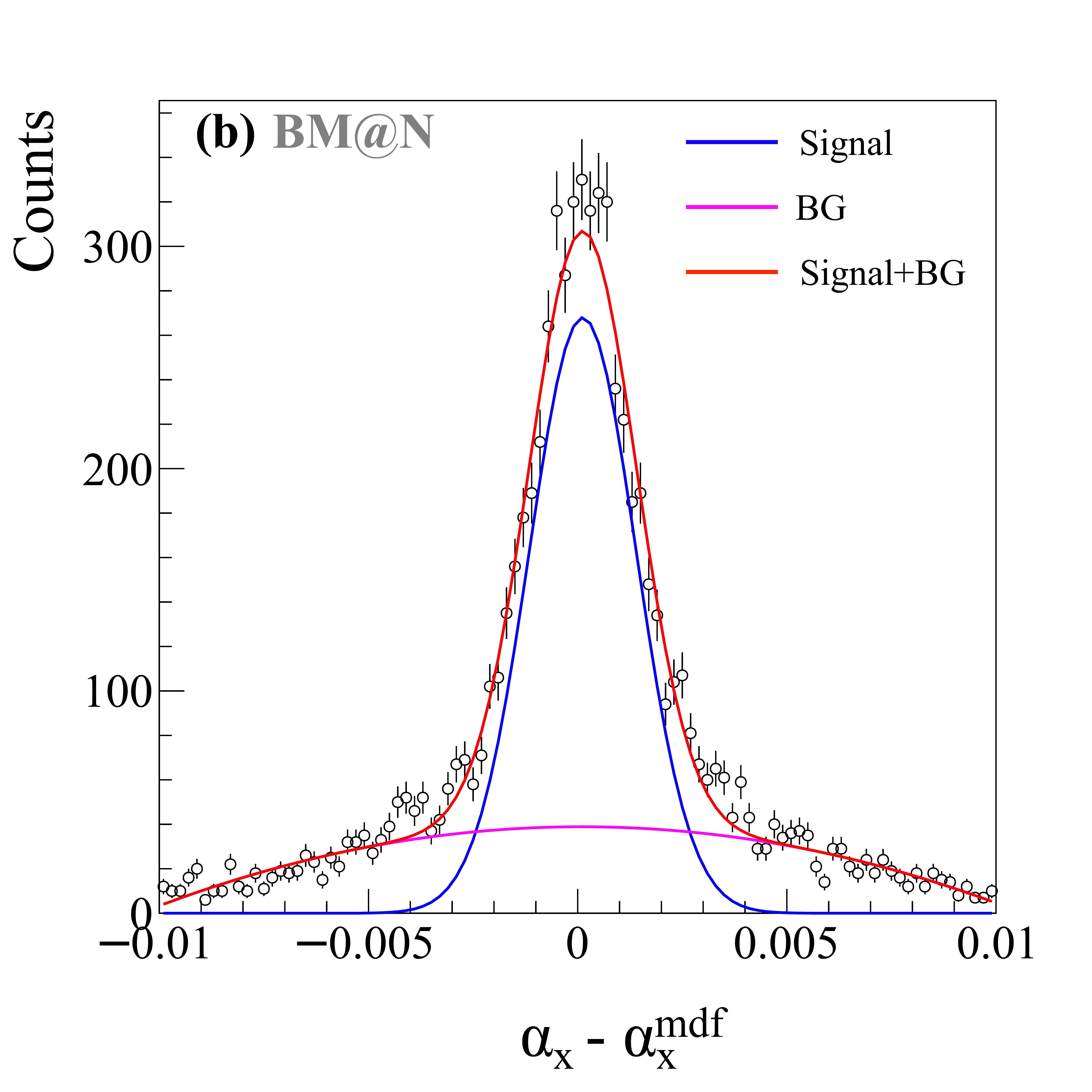}
	\caption{$\vert$ Track matching. {\normalfont (a) Correlation between $\alpha_x$ angle measured upstream of the magnet and the $\alpha_x^{mdf}$ reconstructed by the MDF for unreacted $^{12}$C beam . 
	(b) Residual distribution $\alpha_x^{mdf} - \alpha_x$ fit with a Gaussian peak and wider underlying contribution (``BG'' as second order polynomial).}}
\label{fig:tx_matching}
\end{figure}

\begin{figure}[]
\centering  
	\includegraphics[width=0.4\linewidth]{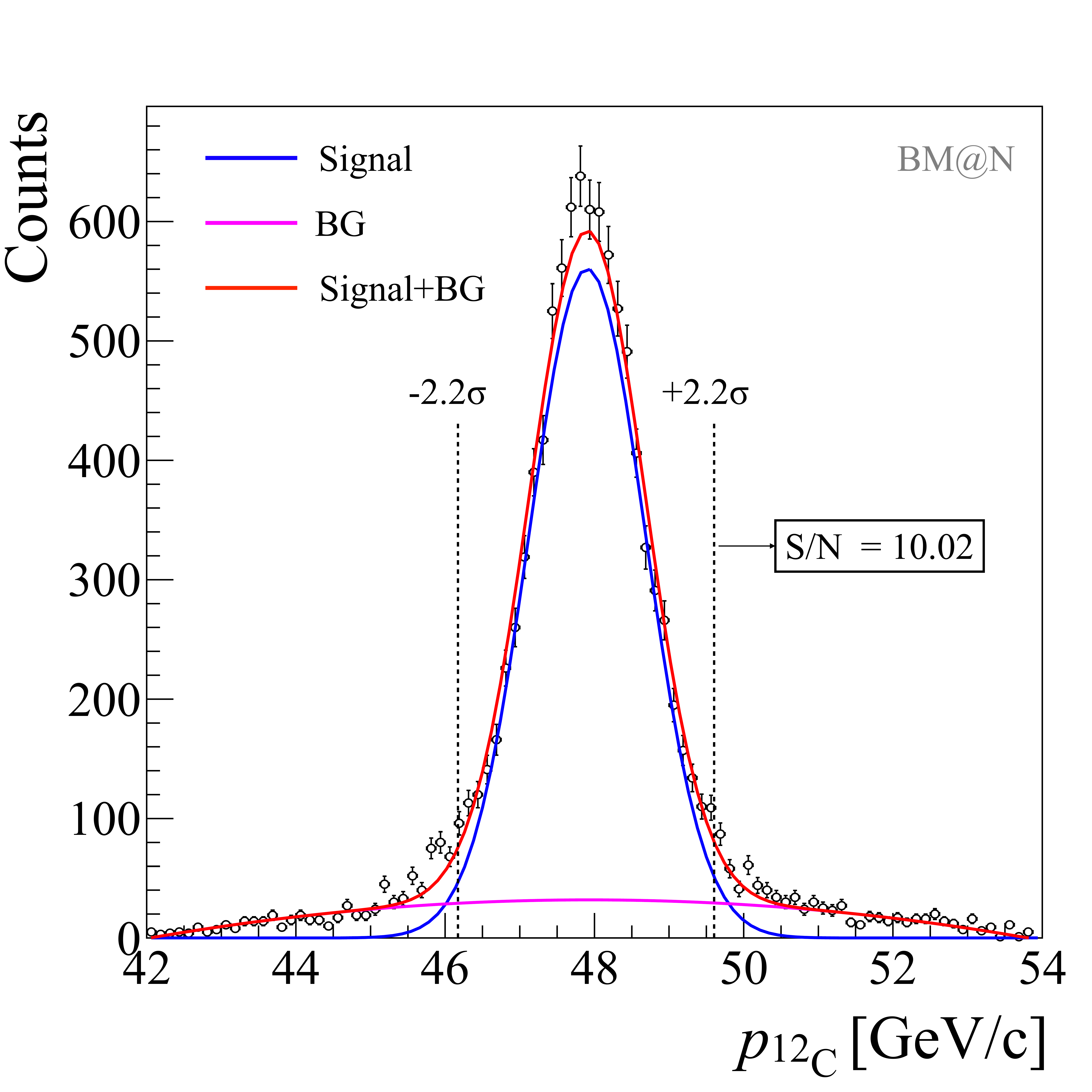}
	\caption{\bf $\vert$ Fragment-momentum resolution.
{\normalfont Total momentum for $^{12}$C measured with empty target, fitted with a Gaussian and possible underlying contribution (``BG''). The signal-to-noise ratio $S/N$ is 10.0.}}
\label{fig:mdf_resol}
\end{figure}

The fragment tracking efficiency is $(39.5^{+1.7}_{-2.6})\%$, obtained for an empty target run and given with respect to the incoming and outgoing $Z=6$ ion.
This tracking efficiency includes the involved detector efficiencies, as well as the reconstruction and matching efficiency of good single tracks.
We define the tracking efficiency for $^{12}{\rm C}$ as ratio of events, incoming carbon $^{12}{\rm C}_{\rm in}$ vs. carbon downstream the target $^{12}{\rm C}_{\rm out}$, with
\begin{equation}
\epsilon_{\rm track} = \frac{\# ^{12}{\rm C}_{\rm out}}{\# ^{12}{\rm C}_{\rm in}}= \frac{\# ({\rm Good \; track)}\& (Z_{\rm in}=6) \& (Z_{\rm eff}=6)}{\#(Z_{\rm in}=6)\& (Z_{\rm eff}=6)}, 
\end{equation}
where a ''good track'' is defined by
\begin{itemize}
 \item Tracks in one of the upstream detector systems and in DCH. 
  \item Exactly one reconstructed matched global track based on the combined information from upstream detectors and DCH as explained above.
  \item A ``good'' $P/Z$ value: for $^{12}{\rm C}_{\rm out}$ the $P/Z$ value is expected to be centered around 7.98~GeV/c (for beam momentum of 47.9~GeV/c), cf. Fig.~\ref{fig:mdf_resol}. 
	The number of $^{12}$C events corresponds to the integral in a $\pm2.2\sigma$ range of $P/Z$, as applied on average for the fragment selection.   
	The uncertainty to the tracking efficiency is determined from a $(2.2\pm0.45)\sigma$ range which reflects the range in $P/Z$ selection for the different fragments of interest.
	In addition, we consider a systematic uncertainty coming from possible remaining wide tails in the $P/Z$ distribution described by a second order polynomial. The signal-to-noise ratio is 10.0. That contribution creates an asymmetric uncertainty in the efficiency, considered on the $2\sigma$ level (cf. Fig.~\ref{fig:mdf_resol}). 
This systematic uncertainty is considered in the same way for the quasielastic event yield, fitting the $P/Z$ for the different charge selections. 

\end{itemize}
Table~\ref{tab:Eff_com} lists the different contributions to the extracted efficiency. 
\begin{table}[h!]
\center
\caption{\label{tab:Eff_com} The different contributions to the tracking efficiency. }
\begin{tabular}{|c|c|}
\hline
Good track & $\epsilon_{\rm track} (\%)$  \\
\hline
\hline
$Z_{\rm in}=6,$ $Z_{\rm eff}=6$  & 100 \\
Upstream track  & 98 \\
DCH track  & 93 \\
Upstream and DCH tracks  & 91 \\
Global track  & 70 \\
Good $P/Z$  & 40 \\
\hline 
\end{tabular}
\end{table}

The tracking efficiency is reduced from 91\% to 70\% due to the MDF algorithm with the applied matching criteria in x angle and a reconstructed single global track. 
That event sample is further cleaned up requiring a single track in the DCH itself, and additional angular matching condition in the y direction (non-bending direction). See discussion above. 
Together with our analysis selection cuts of a good $P/Z$, the efficiency equals 40\%. 
The reaction probability from in-beam material downstream the target was estimated to be smaller 5\% and thus contributes only a small fraction in fragment misidentification.
We estimated the uncertainty for B isotopes and $^{10}$Be identification using the experimental data. 
We looked at the fraction of $^{11,10}$B ($^{10}$Be) from events with $Z_{\mathrm{eff}} = 5$ $(Z_{\mathrm{eff}} = 4)$. 
$Z_{\mathrm{eff}} = 5$ are dominated by $^{11}$B or $^{10}$B. 
We varied the fragment identification cuts to check the sensitivity of this fraction. 
This resulted in a very similar uncertainty as for $^{12}$C, and therefore we adapt the same uncertainty.
$Z_{\mathrm{eff}} = 4$ events are associated with several Be isotopes, or a combination of lighter fragments. 
In this case, to evaluate the uncertainty, we looked at the fraction of $^{10}$Be from events with $Z_{\mathrm{eff}}$ = 4, and changed the identification cuts to evaluate the sensitivity. 
This resulted in $\sim15\%$ difference (as opposed to 5\% for C and B). 
Therefore, for $^{10}$Be, we consider $\epsilon_{\mathrm{track}} = (39.5^{+5.1}_{-7.8})\%$.
For the overall fragment identification efficiency an additional $(83\pm6)\%$ efficiency for the measurement of the outgoing charge in BC3 and BC4 needs to be added.

\bigbreak
{\noindent \bf 3. Reaction-vertex reconstruction} 
The reaction vertex is reconstructed whenever one track is reconstructed in each arm of the TAS.
This requires at least one hit in the GEM and RPC systems to form a linear track in each arm.
We consider only single-track options from the hit combinations.
The coincident two tracks that come closest, formed from all possible hit combinations, determine the vertex position along the beam line in the $z$ direction.  
Alignment procedures within the GEM-RPC system, the left and right arm, as well as relative to the incoming beam are applied.
The initial detector positioning relied on a laser-based measurement, the alignment relative to the other detector systems and the beam using experimental data was done as mentioned before.
The quality of the tracks is selected according to their minimum distance, a selection criteria of better than 4~cm is applied in this analysis. 
Given the smaller angular coverage of the RPC system compared to the GEMs and detector inefficiencies,
the track reconstruction efficiency is 40\%, with an RPC detection efficiency of about 85\%.
\begin{figure}[]
\centering  
	\includegraphics[width=0.4\linewidth]{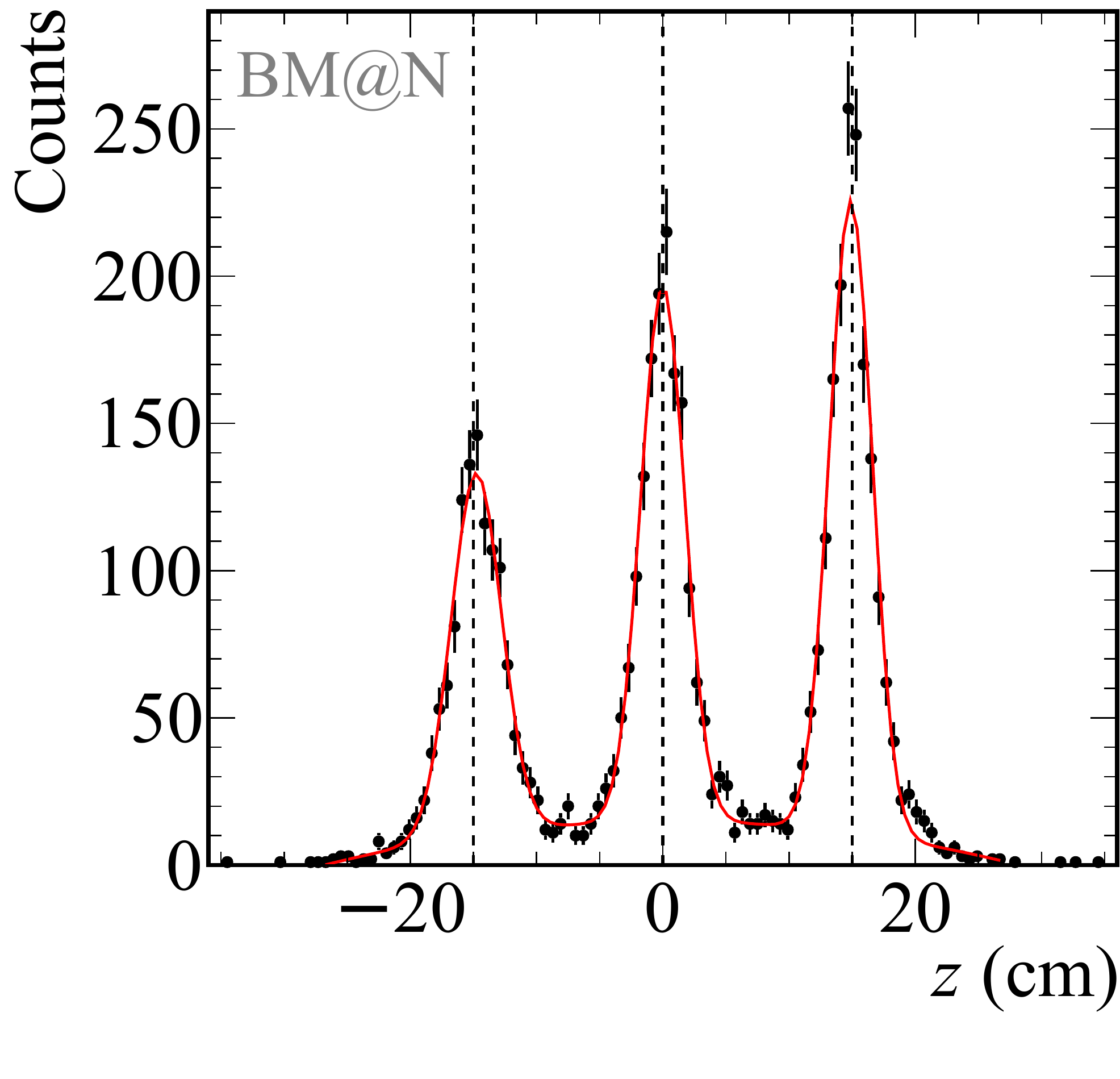}
	\caption{$\vert$ TAS vertex. 
	{\normalfont Vertex in $z$ direction for 3 Pb foils at the target position to determine the position resolution of the vertex reconstruction. 
	The position resolution is 1.8~cm ($1\sigma$), the fit is shown by the red line (plus background). 
	The dashed black lines indicate the absolute position alignment at $z=\pm15$~cm and zero.}}
\label{fig:vtx_3pb}
\end{figure}

The position resolution in $z$ was determined by placing three Pb foils separated by 15~cm at the target position.
The reconstructed vertex position is shown in Fig.~\ref{fig:vtx_3pb}, clearly three distinct peaks at a distance of 15~cm representing the Pb foils are reproduced.
Given the width of each peak, the $z$-position resolution from the two-arm spectrometer is on average 1.8~cm ($1\sigma$).
Knowing the vertex and the track position in the RPC, the flight length is determined.

\begin{figure}[]
\centering  
	\includegraphics[width=\linewidth]{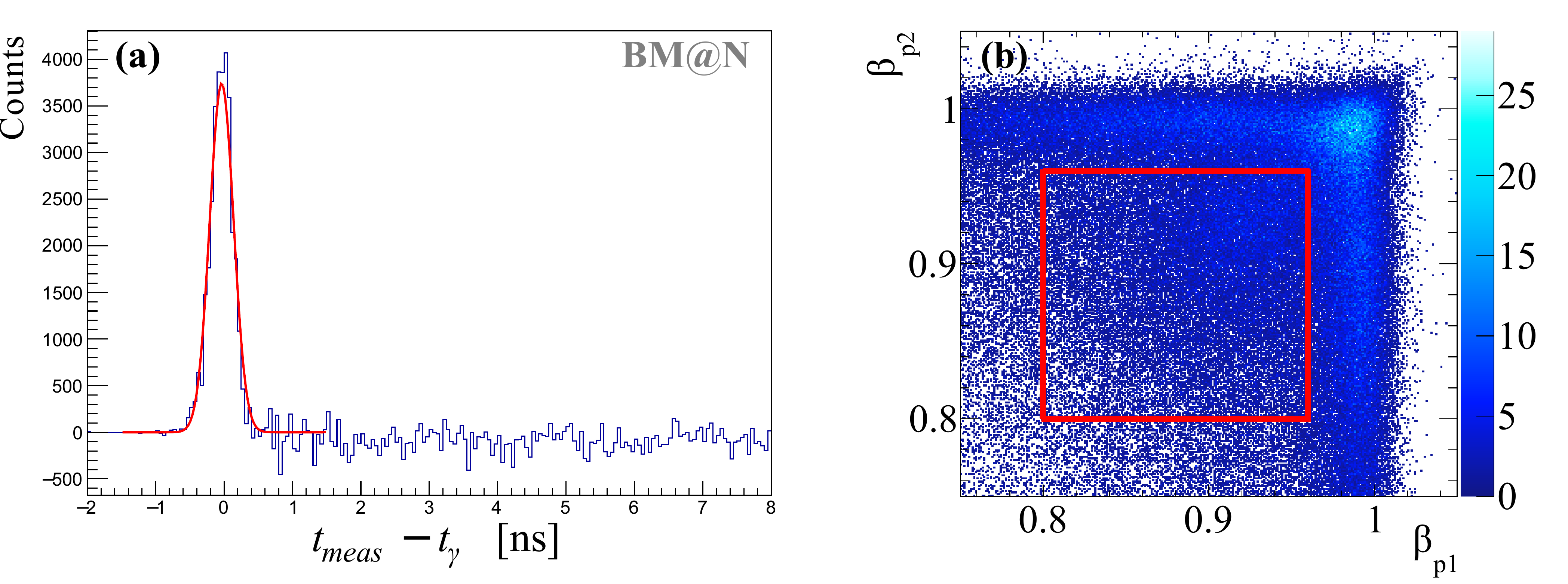}
	\caption{$\vert$ TAS timing. 
	{\normalfont (a) Result of RPC ToF calibration, $\gamma$ peak arising in subtracted spectrum for Pb target runs with and without Pb sheets directly in front of RPC.
	The extracted ToF resolution is 175~ps ($1ß,\sigma$).   
	(b) Basic velocity condition to select protons, the velocity cut in the left and right arm are indicated by the red lines.}}
\label{fig:vtx_mindist}
\end{figure} 

\bigbreak
{\noindent \bf 4. ToF calibration and proton momentum reconstruction resolution.} 
The time-of-flight (ToF) calibration for the RPC is done by measuring gamma rays emitted from interactions with a single-foil Pb target.
A 9~mm thick single Pb target was installed at the center position of the LH$_2$ target. 
In addition, a thin lead sheet was placed directly in front of the RPCs to convert gammas to charged particles.
Measurements were done with and without the  RPC lead sheet and the difference in the measured ToF spectrum for the two measurements was used to isolate gamma rays events.
The subtracted ToF spectrum is shown in Fig.~\ref{fig:vtx_mindist}a,
presenting a total ToF resolution (including the $t_0$ resolution) of 175~ps.
Together with the time-of-flight that is measured between the start counter BC2 and the RPC, the total proton momentum can be determined.
For a 2~GeV/c proton this corresponds to $\Delta\mathrm{ToF}/\mathrm{ToF}\sim0.95\%$ which translates into a total-momentum resolution of 5.3\% in the laboratory system and $\sim60$~MeV/c for the missing momentum from the two protons in the $^{12}$C rest frame.

Fig.~\ref{fig:vtx_mindist}b shows the $\beta$ distribution of measured charged particles in the TAS  with the initial velocity selection cut of $0.8<\beta<0.96$ applied for each particle shown as a red square.

\end{widetext}
}

\clearpage

\end{document}